\documentclass[pdflatex,sn-apa]{sn-jnl}

\usepackage{graphicx}%
\usepackage{multirow}%
\usepackage{amsmath,amssymb,amsfonts}%
\usepackage{amsthm}%
\usepackage{mathrsfs}%
\usepackage[title]{appendix}%
\usepackage{xcolor}%
\usepackage{textcomp}%
\usepackage{manyfoot}%
\usepackage{booktabs}%
\usepackage{algorithm}%
\usepackage{algorithmicx}%
\usepackage{algpseudocode}%
\usepackage{listings}%

\usepackage{booktabs}       
\usepackage{hyperref}
\usepackage{graphicx}
\usepackage{multirow}
\usepackage{tablefootnote}
\usepackage{framed}
\usepackage{csquotes}
\usepackage{enumitem}

\begin{document}

\title{The explanation dialogues: an expert focus study to understand requirements towards explanations within the GDPR}


\author[1,2]{\fnm{Laura} \sur{State}}
\author[3]{\fnm{Alejandra} \sur{Bringas Colmenarejo}}
\author[4]{\fnm{Andrea} \sur{Beretta}}
\author[1]{\fnm{Salvatore} \sur{Ruggieri}}
\author[1]{\fnm{Franco} \sur{Turini}}
\author[3]{\fnm{Stephanie} \sur{Law}}
\affil[1]{\orgdiv{Department of Computer Science}, \orgname{University of Pisa}, \orgaddress{\street{Largo B. Pontecorvo, 3}, \city{Pisa}, \postcode{56127}, \country{Italy}}}
\affil[2]{\orgname{Scuola Normale Superiore}, \orgaddress{\street{Piazza dei Cavalieri, 7}, \city{Pisa}, \postcode{56126}, \country{Italy}}}
\affil[3]{\orgdiv{Southampton Law School}, \orgname{University of Southampton}, \orgaddress{\street{University Road}, \city{Southampton}, \postcode{SO17 1BJ}, \country{United Kingdom}}}
\affil[3]{\orgname{ISTI-CNR}, \orgaddress{\street{Via G. Moruzzi, 1}, \city{Pisa}, \postcode{56124}, \country{Italy}}}

\abstract{

Explainable AI (XAI) provides methods to understand {non-interpretable} machine learning models. 
However, we have little knowledge about what legal experts expect from these explanations, including their legal compliance with, and value against European {Union} legislation.
To close this gap, we present the \textit{Explanation Dialogues}, an expert focus study to uncover the expectations, reasoning, and understanding of legal experts and practitioners towards XAI, with a specific focus on the European General Data Protection Regulation.
The study consists of an online questionnaire and follow-up interviews, and is centered around a use-case in the credit domain.
We extract both a set of hierarchical and interconnected codes using 
grounded theory, and present the standpoints of the participating experts towards XAI. 
We find that the presented explanations are hard to understand and lack information, and discuss issues that can arise from the different interests of the data controller and subject. 
Finally, we present a set of recommendations for developers of XAI methods, and {indications of} legal areas of discussion.
{Among others, recommendations address the presentation, choice, and content of an explanation, technical risks as well as the end-user, while we provide legal pointers to the contestability of explanations, transparency thresholds, intellectual property rights as well as the relationship between involved~parties.}
}


\maketitle

\section{Introduction}

Automated decision-making (ADM) systems are increasingly used in critical domains such as credit, housing, or 
healthcare. 
{However, such}
systems are often based on non-interpretable machine learning (ML) models which have been shown to exhibit algorithmic harms such as {social} bias
\citep{DBLP:journals/widm/NtoutsiFGINVRTP20,ayre2018baked,brozek2023black}. 
Hence, the automatization of high-stakes decision-making processes has {generated} major consequences and effects on individuals' lives, rights and freedoms \citep{ProPublica, AustrianEmployment}.
Furthermore, algorithmic processing tends to be neither transparent nor interpretable~\citep{burrell2016machine}. 
This creates an understandability problem that makes it usually neither possible nor easy to provide an explanation of the system's behavior despite how crucial this information can be, e.g., for the data subject to understand how the specific decision {relating to them} was reached, and to act upon this knowledge.

{One answer to this understandability problem is}
the design and development of understandable algorithmic systems, 
{a key task of the}
field of \textit{eXplainable Artificial Intelligence} (XAI).
{This field offers}
different methods that clarify the functioning of non-interpretable and \textit{obscure} {ADM} systems and provide explanations to individuals so they can understand and act upon the decisions affecting them.{\footnote{Here, we focus on methods that provide an external explanation to ADM systems. Other approaches focus on building ADM systems that are transparent \enquote{from the start}, see e.g.~\citep{DBLP:journals/csur/GuidottiMRTGP19}. However, these systems are beyond the scope of this work.}}
{Such} methods 
are proposed as one solution to detect, understand and possibly prevent the unexpected and undesirable effects that ADM systems can pose and create.
Further, efforts in XAI, mainly driven by scholars in computer science, may be {compliant} with legislative developments towards algorithmic transparency and accountability. Specific examples include the right to information and explanation encompassed in the \textit{European General Data Protection Regulation} (GDPR) \citep{EUdataregulations2018} and recently {explicitly} included in the
\textit{European Artificial Intelligence Act} (AI Act)~\citep{aiact}. 

However, while a significant number of XAI methods (explanations) have already been {advanced}, we know little about the perception, expectations and reasoning of legal experts and practitioners towards these explanations. 
With this work, we are closing this gap: {that is to say, we inquire}
about the legal compliance of XAI methods with respect to explainability and information requirements. 
{Legal compliance is thereby measured against the capability of XAI methods} to make automated decisions understandable and {to verify their} lawfulness and fairness.
We present the \textit{explanation dialogues}, an expert focus study that is designed to reveal the perceptions, expectations, and reasoning of both academics and professionals 
that work on the intersection between AI and law on legal matters and the compliance of AI systems, towards explanations of ADM systems.
The study consists of both an online questionnaire and follow-up interviews where we presented a purposively selected group of experts with different types of XAI explanations.
{These explanations are based on a use-case of an ADM system}
in the credit domain. 
We collected nine valid questionnaire results, conducted six interviews with legal professionals {and drew on} grounded theory, as well as a simple quantitative analysis, {for the data evaluation}.

\paragraph*{Scope}

{In this work, we}
discuss ADM systems, 
particularly fully-automated systems that draw on {non-interpretable} ML. {However, while}
our {use-case}
was constructed over such non-interpretable models, the insights and conclusions reached can be applied to any explanation about automated decisions, regardless of {the model they rely upon.}
Thus, our contributions do not exclude systems that rely on (formally recognized) interpretable models, because these systems may also necessitate an explanation for (lay) end-users as they can be too complex to be understood.

From a legal perspective, 
our study aims to test the transparency and information framework
established by the GDPR~\citep{EUdataregulations2018} towards automated decisions with legal or similarly legal effects on individuals, particularly in Article 22 and its respective Recital 71. The case was designed as a credit application in the context of a contractual relationship between a bank and its customers. Although other {legislative frameworks} would apply in a real-case scenario, for the purpose of this study we limited the scope to the aforementioned {provisions of the} GDPR.
{We acknowledge that the AI Act has introduced a new right to an explanation of individual decision-making on the basis of the output from a high-risk AI system. Its relevance and applicability in terms of information and explanation requirements is undeniable. However, the scope of this paper is limited to the GDPR because at the time of its design, the AI Act was merely a draft that initially included no direct reference to an individual right to explanation. }

{Regarding XAI methods,} this paper is restricted to {methods that can be applied to any model} (model-agnostic), {and explain the whole system (global) or only a single decision (local)}.

Due to the interdisciplinarity of our paper, the intended audience are both legal and technical experts, academics and practitioners who can benefit from our insights about the intersection between what the legislator requires and expects from the explanations of automated decisions and what an XAI explanation can truly offer. Thus, we pay equal attention to {relevant} legal and technical {contexts and spheres of application}.

\paragraph*{Contribution}

With this work, we make the following contributions. 
\textit{First}, building on the questionnaire and interview, and an analysis that is guided by grounded theory, we present our two results specific to the understanding, expectations, and reasoning of the participating legal scholars:
\begin{itemize}
    \item We present a set of hierarchically-ordered and interconnected codes that summarize the findings from our expert focus study. Each of the codes is unique, and presented together with a definition.
    \item We summarize the standpoints of legal experts towards the presented XAI explanations, and classify them via the set of previously-derived codes.
\end{itemize}

While we find that the presented XAI methods face both shortcomings in terms of understandability, presented information, and suitability to exercise {relevant rights with regard to} the data subject and the controller, we also discuss issues that may arise from possibly different interests of the data controller and subject. 
Regarding the legal compliance of the presented explanations, the results {are} mixed, i.e., we cannot state that one presented method complies {with} the GDPR. 
{However, we can assert that the perceived conformity of the explanations to the GDPR is closely connected to how they allow individuals to exercise their rights.}

\textit{Second}, {deriving} from these results, we put forward both a set of recommendations for developers of XAI methods, and {indications of areas requiring further analysis from a legal perspective}.

While the insights we gathered through this study show clear connections to the literature and debates on explainability, our study offers insights into the reasoning of legal experts presented with state-of-the-art XAI methods –- a study design that is, to the best of our knowledge, unique.

\paragraph{Structure}
The paper is structured as follows:
we introduce the legal, technical and methodological background in Section~\ref{sec:background}.
In Section~\ref{sec:the_explanation_dialogues}, we discuss the design and technical details of our study, as well as the steps of our evaluation. The results are presented in Section~\ref{sec:results}, followed by the discussion in Section~\ref{sec:discussion}, the limitations in Section~\ref{sec:limiations} and the conclusion in Section~\ref{sec:conclusion}.
{In the Appendix, we present details of the study (explanations, questionnaire, dataset) as well as further information on the results obtained (coding tables and multiple choice questions).}

\section{Background and Related Work}
\label{sec:background}

\subsection{Legal background}
\label{sec:legal_background}

{In this section, we present the legal bases of the information, explainability and transparency requirements for automated decision-making that motivates this paper  (Section~\ref{sec:gdpr}),
and thereafter, set out the legal context in which our research questions have been developed
(Section~\ref{sec:legal_premises})}. 

\subsubsection{The GDPR and automated decision-making}
\label{sec:gdpr}

{Article 22 of the} GDPR~\citep{EUdataregulations2018} establishes a right not to be subject to a decision based solely on automated processing, including profiling, with legal or significant effects. However, in accordance with Article 22 (2), the right not to be subject to a particular decision does not arise if 
that decision: 
{1)} is necessary for entering into or the performance of a contract; 
{2)} is based on the data subject’s explicit consent; or 
{3)} is authorized by Union or Member State law. 
Furthermore, the Regulation presents a broad framework that provides a {particular} protection for special categories of data.
Article 22 (4) reflects this by stating an additional limitation whereby automated decisions can only be based on special categories of personal data if the data subject has given explicit consent for the processing of special categories of personal data or if there is a substantial national interest proportionate to the aim pursued through such processing. 

Regardless of these exceptions and limitations, the GDPR remains concerned with the gravity of the outcomes that automated decision-making can have{; that is, they must concern decisions having} legal {effects or decisions which significantly affect the individual}. For this reason, 
Article 22 (3) provides for three suitable safeguards that data controllers should implement, at least, to protect data subjects’ rights and interests when one of the exceptions referred in paragraph 2 applies {and which thus entails that} individuals can be subject to automated decision-making. 
{Thus, Article 22 (3) provides individuals with}
1) the right to obtain human intervention on the part of the controller; 
2) {the right} to express one's own point of view; and 
3) {the right} to contest the decision.

Interestingly, the safeguards and protections required by virtue of Article 22 (3) 
are supported and {elaborated upon in} Recital 71, incorporating, for example, a right to challenge rather than contest the automated decision or to obtain an explanation about such a decision. As per Recital 71, these safeguards are~\citep{EUdataregulations2018}:
\begin{itemize}
    \item specific information {{about the processing}} to the data subject,
    \item the right to obtain human intervention,
    \item the right to express his or her point of view,
    \item the right to obtain an explanation of the decision reached,
    \item the right to challenge the decision.
\end{itemize}

These measures aim to protect the data subject's rights by introducing processes through which individuals could systematically verify the accuracy and correctness of automated decisions as well as the relevance of {such a decision}~\citep{Ordinanza21,A29WPGuidelines}. 

The Article 29 Working Party~\citep{A29WPGuidelines} offered a specific -- although not exhaustive -- list of good practice suggestions for data controllers, among which are recommendations for information provisions. 
Annex~1 \textit{Good practice recommendations} presents some suggestions to the data controller for meeting the requirements of the GDPR provisions on profiling and automated decision making. Concerning the right to information, concrete examples are \citep{A29WPGuidelines}:
\begin{itemize}
    \item the categories of data that have been or will be used in the profiling or decision-making process,
    \item why these categories are considered,
    \item how any profile used in the automated decision-making process is built, including any statistics used in the analysis, 
    \item why this profile is relevant to the automated decision-making process, and 
    \item and how it is used for a decision concerning the data subject.
\end{itemize}

Furthermore, the Article 29 Working Party offered a specific example of the information that shall be provided by the controller using credit scoring to assess and reject an individual's loan application; \enquote{[...] if the controller is reliant upon this score it must be able to explain it and the rationale, to the data subject}, \enquote{[[the controller] explains that this process helps them make fair and responsible lending decisions} and \enquote{[the controller] includes information to advise the data subject that the credit scoring methods used are regularly tested to ensure they remain fair, effective and unbiased}~\citep{A29WPGuidelines}. 

Although clarifying the scope of the right to access by the data subject, per Article 15, Recital 63 of the GDPR offers some protection to data controllers concerning the revelation of trade secrets or intellectual property when addressing transparency and information requirements for automated decision (as referred to in Article 22). 
Recital 63 asserts that \enquote{that right [right to access] should not adversely affect the rights and freedoms of others, including trade secret or intellectual property and in particular the copyright protecting the software}~\citep{EUdataregulations2018}. However, controllers shall not rely on this protection to {deny or} refuse the provision of information as clarified by the Article 29 Working Party Guidelines~\citep{A29WPGuidelines}.

\subsubsection{The study's legal premises}
\label{sec:legal_premises}

The imprecise wording of Article 22, together with the multiple and disparate interpretations {of the same article,}
have {generated} a debate about its meaning, enforceability and effectiveness~\citep{binns2021your, sancho2020automated, roig2017safeguards,tosoni2021right,brkan2017ai}. 

Although we will refrain from delving into the literature on Article 22, the motivation of this paper {derives from} the concrete safeguard of \textit{the right to contest an automated decision} \citep{bayamliouglu2022right, kaminski2021right, sarra2020defenceless} and the {controversial} 
\textit{right to an explanation} \citep{edwards2017slave, wachter2017right, DBLP:conf/fat/SelbstP18, kaminski2021right, malgieri2017right, mendoza2017right, DBLP:conf/aaai/PedreschiGGMRT19}. 
{As it is included in the Regulation's Recitals, instead of in its main text and provisions, the right to an explanation has been the subject of intense debate in academia. 
However, the right to an explanation seems to be a prerequisite for the right to contest an automated decision:}
for the purpose of our case-study, we draw from the premise that Article 22 establishes a \textit{transparency scheme} for legally or similarly significant solely automated decisions on the grounds that the effective implementation of Article 22 inevitably requires to make the decision-making process and the reached decision understandable -- {at least to some} extent -- to the data subject \citep{bayamliouglu2022right}.
 
Correspondingly, the {paper} initially assumes the thesis that the right to contest an automated decision \enquote{serves to perfect more substantive rights of fairness and justice and to preserve rule of law values, by correcting errors, preventing or changing unjust outcomes, and enhancing predictability and consistency of decisions} \citep{kaminski2021right}.
Following this assumption, the right to contest as a safeguard to automated decisions would oblige data controllers to create a mechanism within the automated decision-making process that permits data subjects to challenge automated decisions and to receive an adequate response upon that challenge \citep{bayamliouglu2022right}.    

Our study presumes that making an automated decision contestable, therefore, implies some level of transparency and interpretability regarding the particular decision and the automated decision-making around it. In essence, the right to contest would impose the contestability of the particular automated decision and the implementation of individual transparency and process rights allowing the data subject to inspect the adequacy of the decision in regard to the general principles of lawfulness, fairness and transparency in light of the GDPR~\citep{mendoza2017right} and the pertinent sectoral laws affecting it, e.g. contract law, employment law, and anti-discrimination law. Arguably, the extent of the information provided based on such a right needs to be {sufficient so as} to guarantee an effective exercise of the individual’s rights referred to in Article 22(2). In essence, the right to an explanation would not be respected merely by providing as little information as possible about the decision, but the data controller must provide sufficient content for the individual to understand the reasons, normative basis, and logic that led to the final decision. {An alternatively interpretation} would contravene the principle enshrined in Article 5 of the GDPR as it would attempt to elude the data controller's obligations of fairness and transparency during the processing of personal data. 

Some scholars specify that the right to an explanation could be further limited by the transparency requirement encompassed in Article 12, which compels data controllers to make an effort to communicate information in a way understandable to individuals \citep{ananny2018seeing, kaminski2021right, DBLP:conf/fat/SelbstP18}. {From our perspective}, the principle of transparency -- as referred to in Article 12 of the GDPR -- does not only act as a limitation to the potential misconduct of data controller, but reinforces the safeguard rights established in Article 22(2) by ensuring that the information and explanations {with which} data subjects are provided are not only potentially useful for the exercise of their rights but undeniably concise, intelligent, and easily accessible. 

Be that as it may, neither the GDPR nor the Article 29 Working Party Guidelines elaborate on the content or form of the right to contest and right to an explanation, currently limiting it to a standard rather than a set of specific procedural rules to which controllers must adhere~\citep{kaminski2021right}. Nonetheless, this {paper} follows the argument that the rationale behind these two rights can be presumed from the need to ensure that automated decision-making respects the principles of fairness, legality, and transparency encompassed in Article 5 and Article 12 of the GDPR. 

{It is worth stressing that the national laws of EU Member States have approached Article 22 of the GDPR from different perspectives that have narrowed or broadened the interpretation and implementation of the provision and so of the suitable safeguards to automated decision-making. For instance, France and Hungary have explicitly introduced a right to an explanation for automated decisions \citep{HungarianGDRP, FrenchGDRP}, and hence followed a proactive approach to automated decision-making explainability \citep{malgieri2019automated}. This paper, however, does not take into account the particular implementation of the GDPR in each Member State as per their concrete national laws. We strictly follow the wording of the GDPR as enacted by the European Parliament and the Council in its English~translation.}

Although there are undeniably problems and uncertainties with regard to the exercise and implementation of these rights, their existence seems assured. On account of these uncertainties, this paper seeks to identify common expectations from legal experts and practitioners with the aim of clarifying the rules that data controllers shall follow when designing contestable automated decision-making processes.

\subsection{Explainable AI}
\label{sec:explainable_ai}
 
A large number of XAI methods have been proposed in the technical literature. They can be broadly distinguished along two axes~\citep{DBLP:journals/csur/GuidottiMRTGP19,molnar2019}: 
{i)}~whether they are only valid for the data instance in focus {and thus explain a single decision} (\textit{local}), or apply to the full model (\textit{global}); and 
{ii)}~whether they are tailored to a specific model (\textit{model-specific}), or can be used to explain any (\textit{model-agnostic}).

The methods {listed in the following paragraphs} are used in our work. They belong to {the class of} \textit{local}, \textit{model-agnostic} approaches, and are computed 
based on input-output pairs (relative to the model).
{For visual examples of these explanations, we refer to the Appendix~\ref{sec:appendix_explanations}.
}

\paragraph{Feature relevance methods}
Feature relevance methods provide a measure about how \textit{relevant} a feature is for the decision outcome. 
{Therefore, the higher a value assigned by such a method to a feature is, the higher the relevance of the same feature for the decision}.
Methods vary by how this {value} is calculated and therefore by the exact meaning of it. 

In this work, we rely on SHAP (SHapley Additive exPlanations), as {proposed} by~\cite{DBLP:conf/nips/LundbergL17}. 
The method provides positive or negative {relevance} values {of features -- interpreted here as contributions --} towards the prediction of the ML model. 
{These values are additive.}
{Explanations are} either provided for a single data instance (local) or as average over individual SHAP values (global). Further, SHAP is a \textit{model-agnostic} approach.
The calculation of the relevance values is based on a game theoretic approach {(Shapley values)}.

\paragraph{Contrastive explanations}

Contrastive explanations\footnote{Contrastive explanations, {also called counterfactual explanations,} are related to the concept of counterfactuals as understood in the statistical causality literature. However, these concepts are not the same. To avoid confusion, we therefore use \enquote{contrastive}.}
refer to data instances similar to those to explain but with a different prediction outcome.
Such explanations provide an answer to the question \enquote{How should the input data look like, in order to obtain a different output?} which can be easily translated into \enquote{How should the current input change to obtain a different output?}, or to answer \textit{what-if} questions such as \enquote{\textit{What} happens to the output \textit{if} the input changes that way?}.
{In general, the data instance is thereby referring to the individual affected by the decision.}

Contrastive explanations were introduced into the field of XAI by~\cite{wachter2018a}.
We use two methods:
\begin{itemize}
    \item DiCE (DIverse Counterfactual Explanations) is a method that provides contrastive examples in the form of data instances. These are computed based on the minimum distance with respect to the original data instance, but {assuming a different predicted outcome.}
    {Further, the method allows for a \textit{diverse} set of contrastive examples to be obtained. 
    This is considered beneficial to understand a decision, because it provides different explanations of how the outcome of the model can be changed.
    The method was introduced by}~\cite{DBLP:conf/fat/MothilalST20}.
    \item LORE (LOcal Rule-based Explanations) is a method that provides decision rules as explanations. Both a factual rule
    and a contrastive rule {based on a measure of minimal distance}
    {are offered to the user, thus providing both the rationale for the original decision and for obtaining a different outcome.\footnote{The method relies on rules which are a central element of symbolic AI. However, LORE does not allow to reason on the obtained decision rules. We exemplary point to~\citep{DBLP:journals/ia/CalegariCO20} for readers that are interested in symbolic methods for XAI.}}
    The method was introduced by~\cite{Guidotti2022a}.
\end{itemize}

\subsection{Qualitative Research}

We define qualitative research in technology as a methodology that relies on a contextual understanding of different stakeholders (e.g., developers, end-users) when using technology, and with the aim to collect attitudes, behaviors, and insights about those stakeholders to improve the design of future technologies~\citep{blandford2016qualitative}.
Examples of data collection, evaluation and organization approaches in qualitative research are: ethnography, narrative research, 
thematic analysis, and grounded theory~\citep{strauss1990basics}.
What all these methodologies have in common is that they focus on providing an in-depth understanding of the participants about their experiences and perspectives.
Usually, the sample size in such studies is small and purposefully defined on salient criteria (e.g., specific knowledge of the participants), and the data gathering involves close contact between both the researchers and the participants~\citep{strauss1994grounded}.

\subsubsection{User studies in XAI}
\label{sec:user_studies}

A user study is a research method in which the behaviors, preferences, and opinions of the target audience are observed and analyzed. This process typically entails gathering both qualitative and quantitative data through diverse methods like surveys, interviews, and usability testing.
User studies are also an integral part for the evaluation of XAI methods.
They are necessary to comprehend how different types of explanation methods are understood, accepted and used,
and how these methods are subjectively experienced and perceived by different stakeholders. 
Therefore, a user study entails the identification of users' viewpoints, aspirations and expectations, trust, or other qualitative (and also quantitative) measures. 
It also involves the connection of these measures to work practices, as well as legal and political drivers. 
The results of qualitative research in XAI should inform the design and improvement of XAI methods.
However, these results are also important for the design of ADM systems {in general, which must prioritize a human-centered approach to fulfill their intended purpose.}
A key challenge is thereby the involvement of stakeholders into the design of these systems~\citep{DBLP:journals/ijhci/GaribayWAABCFFG23}.

User studies in XAI are still scarce. Considering the case of contrastive explanations, a recent survey found that only $21 \%$ of the approaches are validated with human subject experiments \citep{DBLP:conf/ijcai/KeaneKDS21}. 
A more recent survey on how XAI methods are generally evaluated confirms that also for the whole field of XAI, only $20 \%$ of the methods are evaluated via user studies~\citep{DBLP:journals/csur/NautaTPNPSSKS23}.
{A related work is \citep{DBLP:journals/corr/abs-2210-11584}, a general survey}
on user studies in XAI.
Such studies vary in the task they allocate to the user, used measurements and experimental settings.
To guide the reader through their survey, however, the authors identified five key characteristics of explanations: trust, understanding, fairness, usability, and human-AI team performance. 
These characteristics are based on the identified literature of the study.
Additionally, they provide guidelines for the design of a user~study.

We close this section with the work of~\cite{DBLP:journals/corr/abs-2110-10790}, providing a human-centered perspective of XAI. 
The authors emphasize three points: 
1)~there is no \enquote{one-fits-all} solution in XAI, and specific attention must be drawn to the needs of the end-user;
2)~there is a need for user-studies in XAI; and
3)~theories of human cognition and behavior can inform the design of XAI methods.
Further, the {same} paper {presents} a table that categorizes XAI methods by the question {they are able to answer.
In each row, one question is posed, together with a description of how it can be answered, and examples of XAI methods that provide these answers. 
For example, the question \enquote{How}, referring to the global functioning of the model, can be answered by a description of the general model via \enquote{feature impact, rules or decision trees}, possibly displaying only the most relevant ones. This information can be provided, for example, by partial dependency plots~\citep{molnar2019}.
}
We use this table when selecting the XAI methods for our study, aiming for a set of methods that covers as many questions as possible (see Section~\ref{sec:explanation_cases}).
The table was originally derived via an interview study with XAI designers, discussed in detail in~\cite{DBLP:conf/chi/LiaoGM20}.

\subsubsection{Data collection methods}

\textit{Surveys} are a widely used tool, especially in the Human-Computer Interaction literature, to collect expectations of users and to collect initial data on topics that are broad or complex \citep{Muijs2011}. 
They can be defined as a set of self-administered questions which the participants are asked to {complete}~\citep{DBLP:books/el/LFH2017}. 
Surveys are used to explore areas characterized by a scarcity of knowledge and novelty of topics \citep{babbie2020practice}.
Their main advantage is to collect data promptly and to reach the target population of users. 
The current research mostly relies on a low engagement with the final users, collecting quantitative data that are usually gathered through crowd-sourcing platforms (i.e.,~Prolific, AmazonMTurk) without an in-depth understanding of the user and the complex environment in which the interaction occurs.
{Thus,}
surveys are less effective in acquiring detailed information due to the self-administered nature that does not allow opportunities for follow-up questions when interesting phenomena emerge from the data \citep{DBLP:books/el/LFH2017,DBLP:books/wi/RSP2023}. 
\textit{Questionnaires} are closely related to surveys. While they encompass questions, surveys entail the dissemination of these inquiries to a set of participants.
Put differently, questionnaires serve as instruments for collecting data, whereas surveys constitute a methodology for eliciting information by administering questionnaires.

{The second methodology we apply in this study are \textit{interviews}.}
Interviews delve deeply into the topics under research. Through reflective questioning, the researcher can obtain nuanced data from the participants that would have been lost with surveys. Interviews offer an open-ended and exploratory approach. Due to their flexibility, the presence of predefined questions does not harm the flexibility of this method. The researcher can adapt to the participants' responses or inquire about new lines of interest, allowing opportunistic interviewing to improve understanding.
However, the flexibility of interviews brings challenges, such as managing discussions that may lack boundaries. Interviews are inherently more demanding for the researcher compared to surveys, and they require skills that can be achieved with practice.
The process entails active listening, note-taking, and deciding which topics need further exploration.
This higher effort compared to the surveys is one of the limits of this method.
Interviews are also limited regarding the participants, as each interview demands a significant amount of time during the data collection, preparation, and analysis to discern the significant information from the collected data.

In this paper, we apply a qualitative approach based on an online questionnaire -- an instance of a survey -- and {follow-up} interviews.
To overcome the limits of both methods presented, and to support the validity of our study, we {rely on} triangulation
\citep{jupp2006sage,turnerTriangulationPractice2009,DBLP:books/wi/RSP2023}.
{By triangulation, we refer to an approach that combines several perspectives: different data, researchers, theories, or methods~\citep{DBLP:books/wi/RSP2023,jupp2006sage}. Furthermore, there is a distinction between \enquote{Hard Triangulation, which rigorously tests and challenges findings for validation, and Soft Triangulation, which primarily serves to confirm existing results}~\citep{turnerTriangulationPractice2009}. 
}
{Our approach can be classified as soft triangulation.}
To evaluate the collected data, we relied on grounded theory. We discuss this in the next section.

\subsubsection{Data evaluation: grounded theory}
\label{sec:grounded_theory}

Contrary to what its name might suggest, grounded theory is not a ``theory" of qualitative research. Rather, it is a qualitative research method designed to generate new theories that are rooted in the qualitative data collected during the research process. 
Grounded theory was initially recognized within the realm of social science as the outcome of meticulous examination and analysis of qualitative data \citep{glaser2017discovery}. 
Subsequently, \cite{strauss1990basics} employed the term to denote a data collection and analysis method
which was no longer confined to qualitative data. Consequently, grounded theory represents an approach to theory construction encompassing qualitative 
and quantitative data. 
\cite{strauss1990basics} propose that grounded theory is particularly beneficial for complex phenomena that are not yet thoroughly understood. The method’s adaptability can handle complex data, and its constant cross-referencing facilitates the grounding of theory in the data, thereby revealing previously undiscovered issues. While the type of information used for grounded theory analysis is flexible, there is an emphasized focus on theoretical sampling and contextual factors to enhance the potential for future applicability of the findings. 
\cite{strauss1990basics}
prescribed three coding procedures: open, axial, and selective {which we will describe in detail below. Furthermore, we will discuss the principle of theoretical sensitivity that centrally guides grounded theory}. 

We apply grounded theory as an epistemological framework in this paper. 
We opted for {this} approach given the exploratory nature of our study. 
It furnishes a structured framework for organizing the knowledge derived from the process of data collection.

\paragraph*{Theoretical sensitivity}
Theoretical sensitivity pertains to the researcher's knowledge, expertise, self-awareness, and openness of mind. {In grounded theory, this is essential to creating}
categories and properties, 
to build connections between them and to form hypotheses on the emergent theoretical codes \citep{glaser2017discovery}. 
In other words, theoretical sensitivity can be defined as the ability to abstract concepts from the data and report them according to the researcher's personal experience.
{As a} personal quality of the researcher {it thus} allows {the researcher to perceive} 
the nuanced meanings within the data~\citep{strauss1990basics}.

\paragraph*{Coding in grounded theory}

The coding process serves to distill and categorize data, providing a structured framework that facilitates comparisons with other segments of information. 
The importance of coding {lies in its ability}
to highlight the unfolding of events within the observed scene, {and to} emphasize a focus on the dynamics inherent in the coded data~\citep{charmazConstructingGroundedTheory2012}.
The coding phase consists of a three-step, {iterative} process \citep{glaser1992emergence}, 
and aims to increase the reliability of the analysis from the extensive unstructured data.

\begin{description}
    \item[\normalfont \textit{Open coding}]
    The initial stage of open coding entails the recognition of concepts within the empirical data. The analysis starts with no concepts. As the interpretation advances, analogous ideas are grouped together into categories. The coding is termed ``open" due to the absence of a pre-established set of codes, allowing the researcher to discover these as the analysis progresses. 
    In this initial step, the researcher investigates the data, breaking it into smaller units of analysis {and} analyzes them to identify similarities and differences \citep{glaser1992emergence, strauss1990basics}.
    \item[\normalfont \textit{Axial coding}]
    This coding stage pinpoints the overarching themes, namely the core ideas and occurrences, in conjunction with the conditions and strategies of the participants related to these phenomena, such as causal or intervening conditions. 
    \cite{strauss1990basics} define {axial coding} as a process where data is reassembled in novel ways 
    through establishing relationships between categories. 
    {Thus, building on} 
    the open coding phase, 
    the researcher begins to draw connections between the themes {that emerged previously}
    through a combination of inductive and deductive thinking.
    {Further,} axial coding focuses on specifying categories and sub-categories.
    \item[\normalfont \textit{Selective coding}]
    In the final stage, the analysis is further developed and interpreted.
    Here, core categories, which are the central phenomena that integrate all other categories, are defined, and a 
    descriptive narrative centered around the core categories is 
    {developed.}
    This entire process is iterative, ensuring its validity through continuous comparisons with the raw data, either confirming or challenging conclusions. This ongoing validation process can uncover gaps in the framework that can only be addressed by additional research using theoretical sampling.
    {The process of identifying core categories} includes 
    establishing connections with other categories, validating these associations, and refining the identified categories, allowing for the development of a concise theory \citep{strauss1990basics, glaser1992emergence}.
\end{description}

\subsection{Connecting this study to previous work}

In the previous three {sub-}sections we discussed the legal and the technical background, as well as the methodology we adopt, including a brief overview on user studies.
We close the background section by first discussing two papers that are closely related to ours, discussing commonalities and differences to our work, as well as
the relation between our work and user studies. 

In \citep{DBLP:journals/dgov/HuynhTHSM21}, the authors discuss explanations for ADM systems in the context of a loan application setting.
The work is motivated by the {absence of a detailed} specification in the regulation of the transparency of ADM systems, with a focus on the GDPR (including the understanding of the technical feasibility of such explanations).
While the general topic and motivation is connected to ours, the design and methodological approach of the paper, their understanding of an explanation, as well as their aim, is different.
The study, conducted by an interdisciplinary team of researchers and regulators, and started with two workshops.
{As a result of the workshops, a set of 13 categories of explanations was presented.}
Among others, each {of these} categories contains a (legal) rationale.
The categorization also {provides} the basis for the subsequent development of an explanation. 
This explanation draws from audit trails (or provenance), is formulated in natural text, and addresses eight out of the 13 categories presented.
{Moreover}, the explanation is available as a demonstrator via an online interface.
The explanation, however, has a broader scope compared to explanations understood in the field of XAI (and our work). For example, it includes information about how data was collected.
Also, while we present specific XAI explanations
and collect answers based on them, the development of a \enquote{concrete approach to help data controllers fulfill some of their obligations} is not the aim of our study. Rather, we aim to draw insights on the perception and expectations of legal scholars towards XAI methods.

The second closely related work is \citep{DBLP:conf/ispw/ChazetteKBS22}. The authors present both a literature review and an interview study with industry professionals, on the design of XAI.
While the approach via an interview study is similar to ours, however, the paper has a strong focus on industry, and does not discuss the legal perspective on XAI. 
This is further illustrated by the authors providing some concrete guidelines on the development of XAI systems, formalized into six core activities.
Further, while we provide some recommendations to practitioners in the conclusion of our study, we do not {present} concrete steps {on how to build an} explanation.

Further, our work is closely related to the body of work about user studies, users' interactions and experiences towards XAI (see Section~\ref{sec:user_studies}), 
as well as users' perceptions of justice on automated decisions~\citep{DBLP:conf/chi/BinnsKVLZS18}.
However, our main purpose is distinct.
The crucial difference is that while we do care about the qualitative aspects of explanations, such as comprehensibility,
we do not merely consider them as inherent qualities of the explanations, but as both legal prerequisites for the explanations' validity and legality, and as preconditions for the subsequent effective exercise of related user's rights.
We are therefore analyzing -- via a questionnaire and an interview, and guided by methodology from grounded theory -- experts' experiences with explanations, with a strong focus on understanding how their reasoning and their perceptions could be translated into rules and recommendations that ensure the validity and legality of the explanations.

\section{The Explanation Dialogues}
\label{sec:the_explanation_dialogues}

The explanation dialogues project investigated how XAI methods are perceived and disputed by legal scholars. 
We organized the study in two parts: 1) an \textit{online questionnaire} in which we presented to the participants different types of explanations, and asked them detailed questions about each of the explanations, 
and 2) a \textit{follow-up interview} to better understand answers from the questionnaire.

\subsection{Research questions}
\label{sec:research_questions}

We designed the expert interview study to answer to the following two research questions (RQ):

\begin{description}
    \item[RQ1] How do legal experts reason about explanations for ADM systems, and how do they judge the legal compliance of existing methods?
    Some aspects to consider for a presented explanation:
    \begin{description}
        \item[(a)] Is the explanation complete or incomplete w.r.t.~the expectations of the legal scholars, 
        and is some information given by the XAI more relevant than others?
        \item[(b)] Is the explanation compliant to the GDPR, and is there a preference towards a specific method, or presentation type?
        \item[(c)] Does the legal reasoning change, when presented with the explanation of a true positive/false positive?
    \end{description}
    \item[RQ2] Do legal experts understand and trust explanations for ADM systems, and what are the steps identified to go forward?
    Some aspects to consider are:
    \begin{description}
        \item[(a)] How well are the presented explanations understood?
        \item[(b)] Which gaps in presented explanations are identified? How can the presented explanations be improved? 
    \end{description}
\end{description}

In the Appendix~\ref{sec:appendix_research_questions_design} we present a table that summarizes how the research questions {and subquestion} map with the questionnaire and interview design.
There is some overlap between RQ1 and RQ2 {that should be taken} into account which became even more apparent during the follow-up interviews.
{For the questionnaire, the overlap between RQ1 and RQ2 arises from the mapping between the RQs and the questionnaire: the two RQs are connected via all questions of the questionnaire, excluding question 1 and 2 that are specific to RQ2.}

\subsection{Design details}

In this section, we present details on the design and implementation of both the questionnaire and the interview.
The expert study was tied to a specific \textit{real-world} application scenario: an ADM system that determines whether credit is granted, or not, {where that decision} is provided by a private actor, situated in the EU.
We fixed a \textit{specific scenario}, because we needed to determine the specific norms and laws that apply to the case.
The choice of the credit domain was determined by being a highly-relevant \textit{high-stakes} decision scenario, and 
appearing frequently in the literature on XAI (and the closely related field of Fair-ML)~\citep{DBLP:conf/fat/BordtFRL22,DBLP:journals/corr/abs-2010-04050,DBLP:journals/ethicsit/AlvarezCEFFFGMPLRSSZR24}.
{Why did we} rely on a real-world but hypothetical case, {instead of} observing and analyzing the processes followed by a real bank to provide information and explanation to its customers?
{This was} due to time and resources constraints. {Further, studying the processes followed by a real bank} would have required us to take under consideration too many variants and explanations; {instead, relying on a hypothetical use-case} allowed us to set a quasi-controlled scenario.

\subsubsection{The loan application scenario}
\label{sec:loan_application_scenario}

The loan application scenario involves a bank, an internal consultant of the bank, and a customer who is applying for a credit.
The bank uses an {ADM} system to assess the creditworthiness of the customer.
This system is trained to predict a risk score. 
Credit (or loan) is granted if the score is below the 0.5 threshold, otherwise the credit is denied.
Furthermore, {the bank uses} an XAI method to provide an explanation to the customer about the approval or rejection of the application, as previously determined via the ADM system. 
Therefore, the end-user of the explanation is a lay person (the \enquote{average} bank customer).
The bank is a private actor and thus must comply with the GDPR (see also Section~\ref{sec:legal_background}). 

While the interests of the bank may be different to the interests of the customer, i.e.,~the scenario may be adversarial \citep{DBLP:conf/fat/BordtFRL22}, we assume that the explanations are given out voluntarily by the bank, and are truthful to the underlying model, i.e.,~there is no intentional attempt to manipulate them.
We discuss the limitations of this assumption in Section~\ref{sec:limiations}.

The \textbf{participant of the questionnaire} was explicitly asked to answer the questions about the explanations from the perspective of the internal legal consultant of the bank.
Opposed to this, the role of the participant in the interview was much broader. We asked questions from both a general perspective (e.g., discussing a possible conflict of interest between the bank and the customer), and from a specific one, if needed.

{
The \textbf{ADM system} is a random forest model, trained on a dataset from the South of Germany (see the Appendix~\ref{sec:appendix_dataset}).
The dataset was split with a ration of $8:2$ for training and testing, and the model training resulted in a final accuracy of $0.815$. 
}

\subsubsection{Explanations and cases}
\label{sec:explanation_cases}

\begin{table}[h]
    \small
    \centering
    \begin{tabular}{p{2cm}p{2cm}p{2.5cm}p{2.5cm}p{2cm}}
        \toprule
        \textbf{Method} & \textbf{Type} & \textbf{Presentation} & \textbf{Why?} & \textbf{Question \citep{DBLP:journals/corr/abs-2110-10790}}\\
        \midrule
         - & introduction, general information  & model information, data set and splits, performance and confusion matrix & basic information about ADM system and its use & performance, data, output \\
         Global SHAP & feature relevance & one plot & global information, SOTA method, very often used$^a$
         & How? \\
         Local SHAP & feature relevance & two plots (one per class) (*) & local information, SOTA method, very often used$^a$
         & Why? \\
         DICE & contrastive & two contrastive explanations via table (*)  & local information, SOTA method, legal support for this type of explanation \citep{wachter2018a} & Why not?, How to be that? \\
         LORE & contrastive, rule-based & pair of a factual and a contrastive rule (*) & local information, SOTA method, legal support for this type of explanation \citep{wachter2018a}, understand the significance of rule-based explanations & Why not?, How to be that?, How to still be this? \\
         \bottomrule
    \end{tabular} \caption{Explanations used in the questionnaire: type, method, presentation, reason and answered question according to \citep{DBLP:journals/corr/abs-2110-10790}. All presented explanations were accompanied by a short text about the method and a short textual description. Explanations marked with (*) present the data instance the decision refers to alongside the explanation. \\
    $^a$ 193M downloads of the \textit{shap} python package versus 1.5M downloads of \textit{dice-ml} python package, retrieved on 2024-02-22 from \url{https://www.pepy.tech/}} 
    \label{tab:explanations_why}
\end{table}

We decided to use {four} different \textbf{XAI methods} in our study, based on the following reasons: a) presented information (i.e.,, basic, global or local); b) legal support; c) SOTA (state-of-the-art); d) usage; and e) understanding a specific presentation type. 
In Table \ref{tab:explanations_why}, we display the methods used, including their type, the chosen presentation format, and the reasons behind choosing the method (column \enquote{Why?}). 
We also added a column called \enquote{Question} that lists the specific question that is answered by the method, according to the taxonomy introduced by~\cite{DBLP:journals/corr/abs-2110-10790}. This helps to further delineate the methods, and supports our initial choice of methods.

We presented some general model information, to give to the end-user basic information about the ADM system and its use (model information, the data set and splits, and performance and confusion matrix of the ML model). This answers to questions about \textit{performance}, \textit{data} and \textit{output} \citep{DBLP:journals/corr/abs-2110-10790}. While the first is answered by providing the performance of the ML model, the second is answered by providing information about the data and the third by information about the downstream use of the model.
We further chose to present both global and local SHAP, both feature relevance methods, and present them either as a single plot (global SHAP) or two plots (local SHAP, one per class). SHAP provides (global or local) information about the model, is a state-of-the-art method and frequently used. Global SHAP answer to \textit{How?} while local SHAP answer to \textit{Why?}~\citep{DBLP:journals/corr/abs-2110-10790}.
We also decided to present DICE, a contrastive explanation method, via a table. It provides local information about the model, is a state-of-the-art method, and has some legal support \citep{wachter2018a}. It answer to the questions of \textit{Why not?} and \textit{How to be that?} \citep{DBLP:journals/corr/abs-2110-10790}.
We chose LORE, a contrastive and rule-based explanation method. We presented both a factual and a contrastive rule. The reasons behind choosing this method were the provision of local information, LORE as state-of-the-art method, legal support behind it (as contrastive explanation method \citep{wachter2018a}) and to understand the significance of rule-based explanations. LORE answers to  \textit{Why not?}, \textit{How to be that?} and \textit{How to still be this?} \citep{DBLP:journals/corr/abs-2110-10790}.
All presented explanations were accompanied by a short textual description of the methods in the presentation in the questionnaire.

{In this paper, we use \enquote{XAI method} and \enquote{explanation} and \enquote{XAI explanation} interchangeably. However, one can argue that SHAP is one method that produces two explanations. While we acknowledge this, we refer to SHAP global and local separately to emphasize differences between the global and local explanation.
Furthermore, we present \enquote{general information} in the first row of the table: this information will not be separately discussed and evaluated in the experiments, as it is necessary knowledge to understand the explanations in the first place and thus is given to all participants.}

We presented to the participants the following two \textbf{cases}:
\begin{itemize}
    \item \textit{Case TP (true positive)}: customers that were correctly predicted as \textit{high-risk} by the ADM and therefore are rejected for credit.
    \item \textit{Case FP (false positive)}: customers that were incorrectly predicted as \textit{high-risk} by the ADM and therefore are incorrectly rejected for credit.
\end{itemize}

The {rationale to present these two cases was to understand whether a}
correct or incorrect prediction by the ADM system {would lead to diverging}
answers from the participants 
(see also RQ 1 in Section~\ref{sec:research_questions}).
{Furthermore, since the main focus of this paper is on the data subject and not on the bank, the evaluation centers on the positives instead of the negatives.}

\subsubsection{Participants' selection and validity}
\label{sec:participants_selection_validity}

Since the {paper} aimed to gather the expertise and knowledge of academics and professionals with reputable and renowned careers in legal matters and compliance with AI systems, we used \textit{purposive} sampling, on the criteria that {our participants} are legal experts on the GDPR of the European Union, particularly on explainability and interpretability of ADM systems.
We contacted around thirty {experts,}
which resulted in nine validly filled questionnaires and six follow-up interviews.

While we wanted to include as many experts as possible into this study, a high number of participants was never the first priority. This is for two reasons:
\textit{first}, because we explicitly targeted experts with specific knowledge as previously described, making it -- {given} our time limits and resources -- hard to obtain a high number of participants;
\textit{second}, because from the onset of this study, we do not make claims of general validity but {aim to obtain opinions of selected experts.}

\subsubsection{Online questionnaire}
\label{sec:online_questionnaire}

\begin{figure}
    \centering
    \includegraphics[width=.4\textwidth]{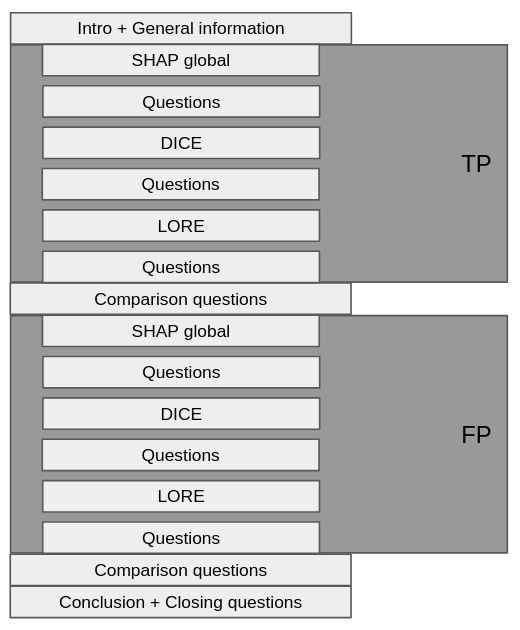}
    \caption{Online questionnaire structure: {each questionnaire consists the Case TP and TN. Further, it contains three different explanation types, one of which is global, two of which are local. Each explanation is followed by a set of questions and each case by a set of comparison questions. The questionnaire starts with an introduction and general information about the ADM system and closes with a conclusion and a set of closing questions}. {The specific} explanations and the order of cases {are sampled randomly.}
    }
    \label{fig:questionnaire}
\end{figure}

In this section, we summarize the structure and randomization approach of the online questionnaire.

\paragraph{Sampling}

Per participant, we randomly sampled a subset of three explanations.
In all cases, we showed global SHAP, and we sampled 2-out-of-3 for the local explanations,
to {be able to compare} different explanation methods between different participants, but avoid an information overload of participants.

{Further, }both Case TP and Case FP were presented to the participants, featuring the previously sampled explanations. Therefore, the \textit{same} explanation method was presented twice to each participant.
This was to compare results within the same participant but between the different cases.

In summary, the participants had to assess $3 \times 2 = 6$ explanations, and we used 
a mix of between- and within-subject (participant) design.
{Regarding the design of the questionnaire, we relied on a combination of between- and within-subject-design.
The first matches (groups of) participants randomly to different cases (here: local explanations) while the second matches all participants with all cases (here: global explanations and cases)~\citep{DBLP:books/wi/RSP2023}.
}

\paragraph{Structure}

The questionnaire can be coarsely described by the following blocks:

\begin{enumerate}
    \item Introduction and general information
    \item Global explanation, followed by questions
    \item Local explanation, followed by questions
    \item Local explanation, followed by questions
    \item Comparison questions
    \item Conclusion and closing questions
\end{enumerate}

Block (2) - (5) was repeated twice, one time per case. The order in which Cases TP and FP were presented to the participant was random. Furthermore, the order of the local explanations (3) and (4) was random, but the same in both cases.
In Figure~\ref{fig:questionnaire}, an example of a sampled questionnaire is shown.
Also, we summarize in Table~\ref{tab:explanations_why} how we present the individual explanations (for example, as plot or text).

We repeated the same block of questions after each presented explanation. The comparison questions {were used to obtain} high-level {insights} that concern all previously presented explanations.
For the evaluation, we {focused on the question blocks after single explanations and the comparison questions.}
We {present} the questions in the Appendix~\ref{sec:appendix_questionnaire}.

\subsubsection{Follow-up interviews}
\label{sec:follow-upinterviewssteps}

The follow-up interviews were designed to {confirm or object to} findings from the questionnaire but also to provide to the participants the possibility to give more detailed answers -- some participants may prefer an interview over a questionnaire, others may have had some doubts about the {questionnaire} that could be expressed and discussed in the interviews.

Therefore, while in the interview we referred back to the loan application scenario (see Section~\ref{sec:loan_application_scenario}), and asked the \textbf{participants} to answer from the previously assigned role of an internal consultant of the bank, the participants also had the additional opportunity to \textit{express their personal opinion} on the presented explanations and during the discussion of the preliminary conclusions.

The construction of the \textbf{interview script} -- closely tied to the questionnaire results -- can be coarsely described by the following steps: 
1) open brainstorming of questions;
2) derivation of preliminary conclusions from the interview\footnote{The preliminary analysis included a screening of the questionnaire data by different researchers and a short discussion. {For the conclusions, see} Section~\ref{sec:questionnaire_summary_final}.}, extension of the {brainstormed} questions, structuring of the questions;
3) refinement and finalization. 
This resulted into an interview structure that had four parts:
\begin{enumerate}
    \item \textit{Introduction}: greeting of the participant, introduction of the project and its objectives, introduction of {interviewers, requesting} for consent for recording and transcription, start of recording and transcribing.
    \item \textit{Warm-up}: asking introductory questions that are not used for the analysis, {for the purpose of making} the participant comfortable with the interview setting.
    \item \textit{Main part}: showing and discussing the preliminary results and asking follow-up questions. This part is structured into two distinct blocks, according to the {preliminary conclusions} (and the RQs).
    \item \textit{Cool-off}: asking summary questions, giving the second interviewer and the participant time to ask questions, close recording and transcribing, asking for feedback, closing.
\end{enumerate}

To remind the interview participant of the explanations, and to support {an understanding of} the preliminary {conclusions},
these conclusions as well as extracts of the explanation were presented on a set of slides. 
The interview was designed to have a \textit{time limit} of 30 minutes.

The interviews were done by two researchers: one had the role of the interviewer while the other had the role of the observer. In the cool-off part, the observer had the possibility to ask a question or share a thought, if wanted.

\subsection{Technical details}

\subsubsection{Explanations and dataset}

To compute the explanations, we used the updated version of the South German Credit Dataset, which is a corrected version of the German Credit Dataset~\citep{Groemping2019}.\footnote{For the data, see \url{https://archive.ics.uci.edu/ml/datasets/South+German+Credit+(UPDATE)}
}
Details about the dataset, dataset splits and necessary manipulations are outlined in the Appendix~\ref{sec:appendix_dataset}.

We presented three local explanations. We computed them for the same data instance, but with {slightly changed} features to avoid a learning effect from previous explanations (see~\cite{DBLP:conf/chi/PaniguttiBGP22}).
We relied on \texttt{sklearn}, the SHAP and the DICE (dice-ml) python packages, and the LORE code from github (lore\_explainer).

\subsubsection{Software}

We used the software \textit{Qualtrics}\footnote{\url{https://www.qualtrics.com/}} to implement the online questionnaire and collect the answers.
The interviews were conducted remotely, and we relied on \textit{Microsoft Teams}. We recorded them as a video, and used the transcription function of \textit{Microsoft Teams} to obtain the interview scripts.

\subsubsection{Anonymization and data storage}

While we asked some information about the background of the participant in the closing questions of the \textbf{questionnaire}, we did not collect any identifying information; thus the questionnaire was fully anonymous.
{However,} the original video data and the transcripts of the \textbf{interviews} were not anonymous. We anonymized the transcripts by replacing any identifying information (names of participants and interviewers, names of institutions) {with non-identifiable tokens (e.g., \enquote{Expert 1} to refer to the first expert participating in an interview).}

\subsection{Evaluation details}

\subsubsection{Data cleaning}

Some steps were necessary to clean the questionnaire data. 
For the interview analysis, we used the transcripts in combination with the videos (if the transcripts were faulty). 
After the cleaning, we had nine questionnaires answered. 
By explanation type, this means nine answers {with regard to}~global SHAP, six answers {with regard to}~DICE, eight answers {with regard to}~local SHAP and four answers {with regard to}~LORE.
The six interviews each had an approximate length of 30 minutes.

\subsubsection{Terminology}

\begin{table}[t]\small
    \centering
    \begin{tabular}{p{3cm}p{8cm}}
        \toprule
        \textbf{Term} & \textbf{Description} \\
        \midrule
        code/topic & A concept extracted from the data via a coding step, each code has a definition (listed in the questionnaire/aggregation coding table). \\[2ex]
        hierarchy of codes & Ordered summary of the extracted codes/the diagram we developed. \\[2ex]
        core phenomenon & \multirow{5}{*}{\parbox{7cm}{The five terms we use to refer to codes in the hierarchy of codes we developed. From high/broad (core phenomenon) to low/narrow (sub-category).}} \\
        theme & \\
        sub-theme & \\
        category & \\
        sub-category & \\
        \bottomrule
    \end{tabular}
    \caption{Overview of key terminology used in the paper.}
    \label{tab:terminology}
\end{table}

The paper draws on terminology from qualitative research. 
Table \ref{tab:terminology} lists the relevant terminology used in the analysis and subsequent sections, with a specific focus on terms used to describe the extracted codes and their relations.
If applicable, we further use the terms \textit{axis} or \textit{spectrum} to describe the different degrees to which we identify a standpoint towards a specific code.

\subsubsection{Steps for the analysis}

\begin{figure}
    \centering
    \includegraphics[width=.9\textwidth]{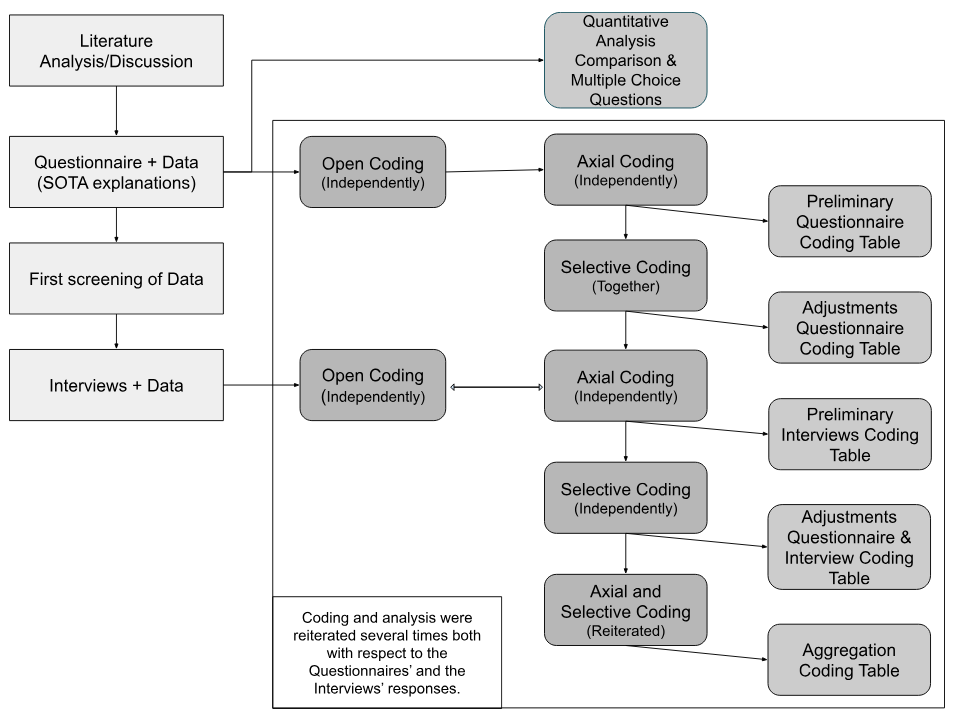}
    \caption{Main steps in the evaluation process.}
    \label{fig:evaluation}
\end{figure}

Our evaluation builds on grounded theory as introduced in Section~\ref{sec:grounded_theory}. Here, we summarize the main steps of the evaluation, taking {also} into account that the evaluation was done by two independent researchers.
{The steps are} displayed in Figure~\ref{fig:evaluation}.

As we use a mix of free text and multiple choice questions, the evaluation consisted both of a qualitative part (based on grounded theory), as well as a statistical part.

\paragraph{Questionnaire/qualitatively}

Regarding the qualitative evaluation, our analysis can be summarized as follows:
\begin{enumerate}
    \item Independently, the two researchers performed \textit{open coding} for all {data of} the questionnaire.
    Rather than per question, codes were recognized per sentence as the researchers acknowledged that full answers could hold a lot of text and therefore meaning. Simultaneously, interesting quotes were highlighted in a separate file, indicating the question and (anonymised) answers {to which} they belong.
    No pre-established set of codes existed before this step, thus, researchers were fully allowed to discover multiple and different codes per similar phenomenon. 
    \item Still independently, the same researchers performed \textit{axial coding} per question, organizing open codes into themes and including as many themes as necessary with their respective axial points.    
    \item One of the researchers developed a table (\textit{Preliminary Questionnaire Coding Table}) with the potential phenomenons and themes {arising from} the questionnaire responses, including their respective axial points. 
    \item Together, both researchers discussed the identified potential themes and performed \textit{selective coding}. Both researchers agreed on six core phenomena which integrated all other themes and initial codes recognized in the questionnaire {data}. These core phenomena {represented} the principal narratives founded in the {data} and {pointed to} single themes as the central axes of a spectrum where the majority of the answers were located.
    \item Subsequently, the \textit{Preliminary Questionnaire Coding Table} was adjusted according to the compromise reached in the previous step, reviewing and revising phenomenons by merging, rearranging and removing.
\end{enumerate}

\paragraph{Questionnaire: aggregation}

The final aggregation, for the qualitative 
answers of the questionnaire, was done over Case TP and Case FP. This resulted from the small difference in the answers to the two cases, and a small number of participants.
We discuss Case TP versus Case FP also in the results, in Section~\ref{sec:questionnaire_summary_of_results}.

\paragraph{Questionnaire: comparison questions}

The comparison questions were evaluated separately from the other open questions, {and without any aggregation.}
We provide a summary of these responses in Section~\ref{sec:questionnaire_comparison_questions_results}. 

\paragraph{Questionnaire/quantitatively}

The evaluation of the multiple choice questions was done via a Python script. We performed a simple cleaning step to {filter} out not admissible answers.
We computed the histograms over single questions and per presented explanation, and used answers to both Case TP and Case FP.
This was done for the same reason as the final aggregation of the qualitative data over Case TP and Case FP.
Therefore, assuming that there are no faulty or missing answers, every participant answered twice to each multiple choice question.
For better comparability, we plotted the different explanation types into one histogram. We display two plots in Figure~\ref{fig:mc_0}. The remaining plots can be found in the Appendix~\ref{sec:appendix_mc_questions}.

\paragraph{Interview}

The qualitative evaluation {was conducted as follows}: 
\begin{enumerate}
        \item The two researchers independently performed \textit{open coding} for the {transcripts of the interviews.}
        \item {Each researcher in their respective coding documents designated a row per question or group of principal and follow-up questions pertaining to the same topic. Then, the interviewees' responses were coded per line or sentence.} 
        Each researcher assigned a code per sentence or idea offered by the interviewees when answering each question. As interviewees were encouraged to answer questions with as much detail and content as they find appropriate, answers usually encompass more than a sentence or idea. Therefore, multiple codes were identified per each interview question and interesting quotes were written down indicating the question and answer they correspond to, so they could easily be traced back. {Notwithstanding that the \textit{Preliminary Questionnaire Coding Table} had already been developed, the researchers were asked to identify codes without relying on or checking the previously agreed core phenomena. For this reason, we still consider this stage of the interviews' analysis as open coding.}
        \item In a second step, both researchers performed \textit{axial coding}, organizing codes into themes, including as many themes as necessary with their respective axial points. 
        \item Subsequently, the researchers identified the main underlying potential themes, {and built} a new coding table (\textit{Preliminary Interviews Coding Table}) {which includes} the selected phenomena and themes recognized in the interview responses. 
        \item {Together, the researcher compared the questionnaire's core phenomena -- included in the \textit{Preliminary Questionnaire Coding Table}- with the themes recognized in the interviews' axial coding. They examined possible overlaps, conflicts, and lacuna in a first round of \textit{selective coding}}
        \item {In a second -and final- round of \textit{selective coding}, the researchers took the \textit{Preliminary Interviews Coding Table} and adjusted it to the finding of the previous steps. They reviewed and revised the themes by merging, rearranging and removing. }
\end{enumerate}

\paragraph{Aggregation}

\textit{Axial and selective coding} were reiterated several times both with respect to the questionnaire and interview responses. Notwithstanding, researchers redid the analysis with special attention to the interview responses; the recognized core-phenomena and their sub-themes were repeatedly compared and validated with the data obtained during the follow-up interviews, with the intention of finding gaps in the framework. 

The evaluation presented in this section resulted in an \textit{Aggregation Coding Table} which reveals a theoretical narrative organized in a hierarchical structure, made up of three core phenomena and their pertinent sub-categories. The table can be found in the Appendix \ref{sec:appendix_coding_tables}.
        
\section{Results}
\label{sec:results}

In this section, we will offer a detailed explanation of the core phenomena recognized both in the questionnaires and follow-up interview responses that led to our theoretical narrative. 

\subsection{Questionnaires}

\subsubsection{Summary of the coding table}
\label{sec:questionnaire_summary_coding_table}

The full coding table of the questionnaire can be found in the Appendix \ref{sec:appendix_coding_tables}.
The six core phenomena that we ultimately distinguished in the questionnaire answers are the following.\\

\noindent\textbf{Relevance of explanation to the overall understanding:} {referring to the contribution of the explanation to the} individual's general understanding of the automated decision and {its} particular relevant details. This phenomenon {has three axes}: an entire contribution, limited or partial contribution, and insufficient contribution. The \textit{understandability for the average consumer} {is a theme of this phenomenon, which indicates} the capacity of the average consumer (reasonably well-informed and reasonably observant and circumspect individual) \citep{Directive2005/29}  to process and understand the explanation/information provided. The axis for this theme are the two sides of the same coin. On the one hand, the possibility or difficulty to understand the decision solely based on the explanation provided. On the other hand, the intention of the explanation's provider to offer confusing or clear information to the recipient so it allows or complicates the decision's understanding.\\

\noindent\textbf{Appropriateness of the delivery method and format:}~ {referring to }the level of suitability of the explanation's format and selected (XAI) methodology. {One of the axis of this phenomenon is }the possibility of the delivery method and format to be confusing or clear. {The other axis is } the intention of the provider to use misleading or helpful methods and formats.\\

\noindent\textbf{Suitability for exercising rights:}~{indicating} the degree of appropriateness of the explanations/informa\-tion in the event that the individual {decides} to act upon the decision affecting them and exercise any of the rights at their disposal. {We identified the axis of "adequate, partially adequate, or inadequately" in regard to the relevance and appropriateness of the explanation to allow individuals to exercise their rights}\\

\noindent\textbf{Significance of the decision's outcome in terms of information requirements:}~{pointing out to} whether the impact of the decision determines the threshold and amount of information requested by the transparency requirements. {The respective axis of this phenomenon are} the demand or lack thereof to adapt the information provided in the explanation on whether the outcome positively or negatively affects an individual.\\
 
\noindent\textbf{Completeness of information:}~{indicating} the amount of information provided.{We identified its axis} in the adequateness or lack of information.\\

\noindent\textbf{Usefulness and necessity of information based on typology:}~{referring to} the requirement and necessity to provide specific and particular information about the decision. Being its axis the need to provide global, local or a combination of both types of information in the explanation. 

\subsubsection{Summary of the results (qualitative)}
\label{sec:questionnaire_summary_of_results}

The particular axis where participants' answers were located depends on the typology of the explanation provided. Therefore, we summarize the results by explanation type.

\paragraph*{SHAP explanation (global)} 

For our participants, the global explanations presented did not contribute to the understanding of the explanation. Global explanations were also considered lacking in information in general. For example, one of our participants highlighted the necessity to include information regarding \enquote{the data that train the model, [or] the final percentage [for being considered high-risk]}. 
\footnote{Throughout the paper, we do not correct errors in the quotes of the participants.}
{Furthermore,} special attention was made to the lack of information regarding  \enquote{reference [to the customer] case rather than a global description}. Particularly, participants highlighted the need to include \enquote{any information pertaining to the individual} such as \enquote{the characteristics of the single features},\enquote{the variable weight on the final result} or \enquote{factors that actually apply to the case}. The plot provided through this method was {also} considered {to be} misleading. In the words of one of our participants \enquote{[...] I have a hunch that the plot is misleading in that a transfer of balance from savings to checking does not increase the chance of credit acceptance}. Global explanations were described as being difficult to understand by an average consumer and not appropriate to exercise individuals rights. {Some of the reflections of our participants in this regard were that} \enquote{as a user, I would need additional training to understand and interpret the explanation}, {or that} \enquote{in trying to build a case for why an individual is creditworthy [...] it is unclear to me how this explanation helped. It does not seem possible to build up a coherent argument solely on this explanation}.
    
\paragraph*{DICE explanation} 

{Such explanations} were considered to contribute to the understanding of the decision and were appreciated due to their provision of information regarding the single features defining the final decision. However, participants were conflicted on whether contrastive explanations  were more or less easy to understand, e.g. \enquote{generally I can understand well the features because I have much more information regarding them but at the same time I have difficulty to understand the explanation model itself}. 
{The} participants agreed on the benefits of explaining in more detail what a contrastive explanation is and how the provided ones were selected. They also acknowledged the positive side of pairing them with a narrative box. That being said, the participants found the actionability of DICE explanations controversial as \enquote{the consumer is empowered to verify that their data are entered correctly}, but \enquote{non-actionable counterfactual [contrastive explanation] do not adhere to my intuition of what an explanation is. There is information there that allows the customer to gain insight into the algorithm, but not a tremendous amount}. A participant further unfolded {that} \enquote{the given explanation seems great to make a customer happy; they are informed about how they can improve their creditworthiness, and can re-apply having improved these factors. While I find it hard to imagine ways to enable an individual to assess discrimination risk in any productive way, at least this explanation didn't help}.

\paragraph*{SHAP explanations (local)} 

Participants expressed their doubt about their relevance and their partially or limited usefulness {to the understanding of the decision}, e.g.~the SHAP plot seems \enquote{somewhat useful}, \enquote{uninformative} or merely \enquote{too difficult for me}. Some participants conceded in the overall understanding of the model itself, but requested more information regarding the single features affecting it.  Participants also differed on {how appropriate the method is}; a participant affirmed that SHAP explanations allow the understanding of the model, whereas two other participants perceived the explanations as \enquote{possible cognitively misleading}. Furthermore, the format of SHAP explanations, concretely its graphic design and plot, was perceived as confusing and not easy to understand. For example, a participant recommended more detail in the explanation or the use of examples as \enquote{I usually understand more written text better than plots   and schedules}. Furthermore, SHAP explanations were perceived as not entirely nor directly understandable for the average consumer since \enquote{a plain written text instead of the plot would probably be more intelligible [for an average consumer]}. Finally, SHAP explanations were described as partially suitable for customers to contest the decision; \enquote{they [customer] know where to further inquire}, admitting that \enquote{they can contest the decision even if they do not understand the explanation}. 

\paragraph*{LORE explanations}

{LORE explanations} were described {quite differently among our participants}; {[they were considered]} \enquote{hardly helpful} but also \enquote{fairly intelligible} and \enquote{more clear and understandable [than the other type of explanations presented to this participant]}. All participants, nonetheless, agree on the lack of {clarity in their delivery format}. Particularly, {they} highlight{ed} the lack {of clarity} regarding how a change in the individual circumstances besides the one suggested by LORE would change the benchmark of the decision. Participants also {considered} how the average consumer would {face challenges in understanding} LORE explanations, e.g.~\enquote{this is too difficult for a random bank consumer}. All in all, {including \enquote{text in natural language} was suggested as an option to improve the intelligibility of the explanation}.  

\paragraph*{Significance} One factor to note in the analysis of our questionnaire is that, with only a couple of exceptions,
participants were indifferent as to whether the decision affects positively or negatively the individual, underlining the importance of how the model functions not on its negative or positive outcome, e.g.~\enquote{if the outcome was true positive, I need to know the same information. It's a matter of model, not the result. I need to know how the model works, doesn't matter strictly the outcome itself}. That being said, one of our participants clarified that the significance of the decision's outcome \enquote{would only change since contesting a positive decision seems unlikely} in the sense that, as another participant pointed out, \enquote{in case of a positive assessment, I would be less demanding}. We cannot establish whether the correctness of the decision was relevant for the participants' assessment of the significance of the decision and, thus, the significance of the decision's outcome.

\paragraph{Case TP vs FP} It is worth recalling that Case TP presents participants with rejection of a loan application based on a correctly predicted high-risk status, whereas Case FP presents a rejected loan application based on an incorrect prediction of the individual as high-risk. This difference in the correctness of the outcome and, therefore, subsequent ADM decision did not, to our understanding, have any special impact on the answers of our participant. In other words, the correctness of the outcomes did not seem to be a consequential element for our participants, which justified why we aggregated the answers of Case TP and FP for each question.  

\subsubsection{Summary of comparison questions (qualitative)}
\label{sec:questionnaire_comparison_questions_results}

The following are {the} conclusions we draw from the {participants'} answers {to the} questionnaire's comparison questions. {All of them} refer to the core phenomenon of \textit{usefulness and necessity of information based on typology}.

\paragraph*{Did you prefer one/several explanations over other explanations -- for example global, local, or specific methods?}

Respondents coincided in their preference of local over global explanations, accentuating their higher level of clearness and intuitiveness in contrast with the uninformative nature of the latter. However, only one respondent assessed a specific method, contrastive explanations -- called \enquote{counterfactuals} by the participant -- on the basis of their \enquote{actionability and usefulness for the customer}. 

\paragraph*{If you’d be free to mix the explanations that were shown to you as you’d wish (type of explanation/format and presentation of the explanation, etc.), how would you do that?}
{We obtained diverse responses in this regard.} One of the respondents recommended paying attention to the interpretability of the explanations suggesting that explainability efficiency is connected to the use of limited methods. Still, {another} respondent proposed to provide as much information as possible in the explanations, although the respondent acknowledged the possible interpretability trade-off that could result from that request. {Including appropriate documentation and combining global and local explanations were also options proposed by other respondents; another participant suggested to rather only present textual and local base explanations.}

\subsubsection{Summary of the multiple choice questions' results (quantitative analysis)}
\label{sec:questionnaire_mc_results}

\begin{figure}
    \centering
    \includegraphics[width=0.475\textwidth]{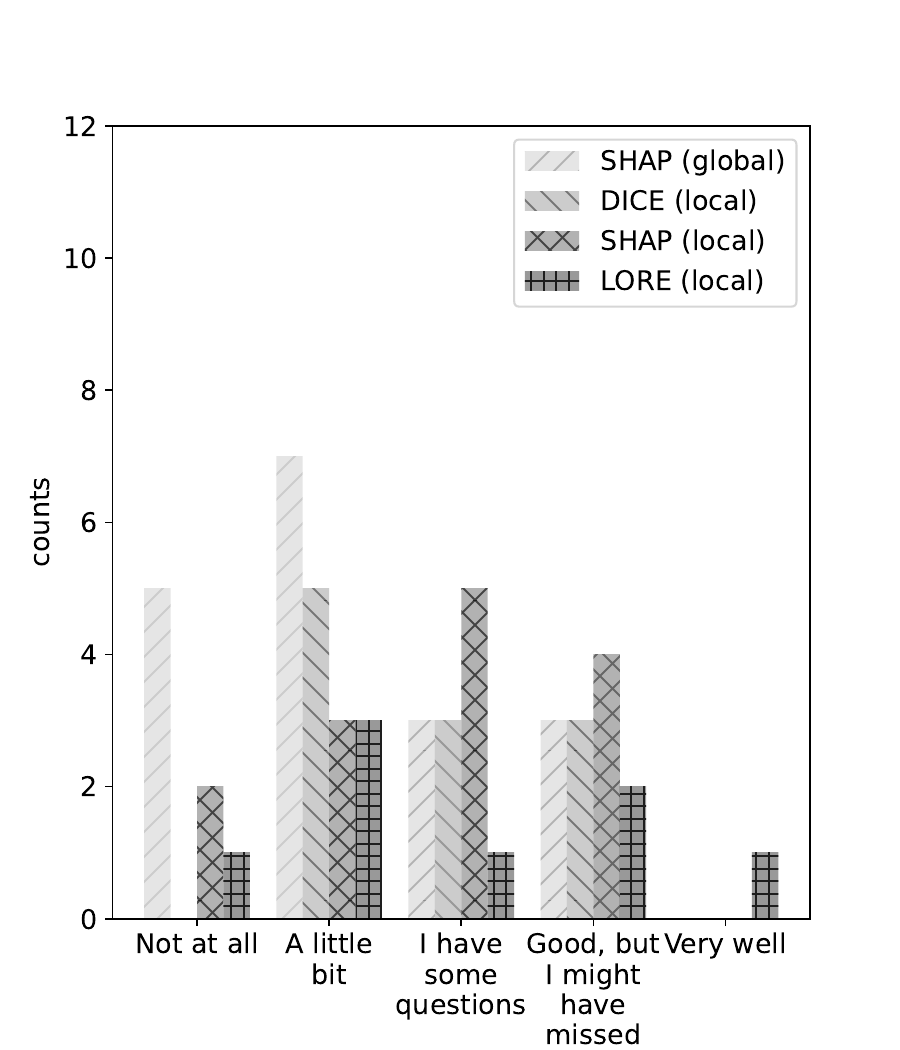}
    \includegraphics[width=0.475\textwidth]{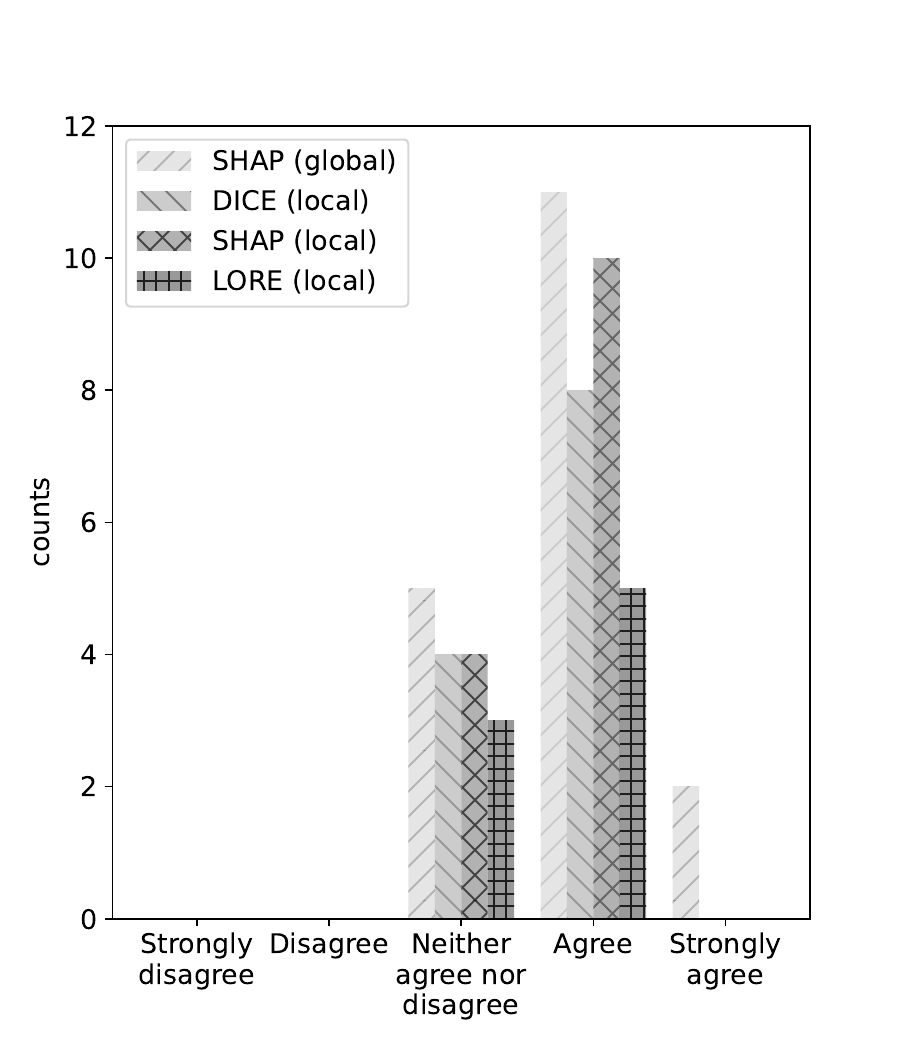}
    \caption{Exemplary results from the multiple choice questions. Left: answers to \enquote{Judge how well you understand the decision based on the shown explanation on the following scale}. Right: answers to \enquote{The explanation provided does not pose a [potential] conflict with the interest and rights of the bank (e.g., intellectual property)}.}
    \label{fig:mc_0}
\end{figure}

Finally, we presented respondents with six statements concerning the explanations they were shown.

Participants were initially asked to \enquote{\textit{judge how well you understand the decision based on the shown explanation}} in a five degrees scale from \enquote{not at all} to \enquote{very well}.\footnote{The degrees being "Not at all", "A little bit", "I have some questions", "Good, but I might have missed something", and "Very well".} From the responses obtained for this particular statement, we can gather some general conclusions. \textbf{Global SHAP} was considered by some of the participants completely unhelpful to understand the decision. {Those who acknowledged the helpfulness of Global SHAP -- to some degree or another --, also recognized the possibility of some information being missing}.
{Participants stated that \textbf{DICE} explanations facilitated their understanding of the decision, but they also acknowledged that they still had certain questions.}
Similarly, most of the participants deemed \textbf{Local SHAP} helpful to different degrees despite some remaining open questions. 
\textbf{LORE} {were the} explanations that received the most disparate evaluations, with participants finding them to be not at all helpful to very helpful passing by all of the other different options. 

To {obtain a more in-depth understanding of the participants’ perceptions} towards the explanation provided, we also asked them to express their level of agreement towards the other five statements. A scale of five different levels of agreement accompanied each statement starting from \textit{I do not agree} and ending in \textit{I totally agree}.\footnote{The degrees being "Strongly agree", "Disagree", "Neither agree nor disagree", "Agree", and "Strongly agree".} Respondents were asked to select among the five options the one {that} best expressed themselves. From the answers received we can draw the following conclusions. 

\paragraph{Global SHAP explanations} They were generally considerd to be unclear and non-understandable for an ``average individual". In particular, this type of explanation was deemed insufficient in providing enough information to data subjects for them to understand the reasons and motives behind the particular decision affecting them. Likely based on this same basis, Global SHAP explanations were not found adequate to allow data subjects to verify the lawfulness and fairness of the automated decision affecting them nor to effectively exercise their right to contest [if deemed appropriate].  

\paragraph{DICE explanations} {These explanations} were more heterogeneously perceived by our participants. For the most part, DICE explanations were regarded as unclear and non-understandable for the average consumer, although no compromise was reached with regard to their adequacy to provide sufficient information for the data subjects to understand the reasons and motives of the automated decisions. Additionally, DICE explanations were, by a small majority, deemed suitable for data subjects to exercise their right to contest, although they were considered inadequate {in terms of verifying} the lawfulness and fairness of decisions affecting individuals.  

\paragraph{Local SHAP explanations} {Such explanations} were perceived with a high level of neutrality, in the sense that a large number of participants neither agreed nor disagreed with the statements provided with regard to this type of explanation. That being considered, participants almost equally agreed and disagreed (or strongly disagreed) on the clarity and understandability of local SHAP explanations. These types of explanations were, by a small majority, reported insufficient in providing information about the reason and motives of the automated decision, although they were equally considered to allow and disallow the data subject to effectively exercise her right to contest such automated decisions. However, they were strongly perceived as inadequate to verify the lawfulness and fairness of the automated decision in regard to other sectoral laws applicable to the case. 

\paragraph{LORE explanations} {Such explanations} were considered to be both clear and understandable for the average consumer and the opposite in equal parts. However, they were considered insufficient to provide information on the motives and reasons for the automated decisions. An equal number of participants either disagreed on the adequacy of LORE explanations to allow data subjects the exercise of their right to contest, or were hesitant about their suitability [nor agreeing nor disagreeing with the pertinent statement]. Opinions were varied regarding whether LORE explanations would allow the verification of automated decisions' lawfulness and fairness. 

\paragraph{Conflict of interests}
Notwithstanding shortcomings and deficiencies identified in the explanations, our participants did not indicate that any of the explanations provided would give rise to a potential conflict of interest [and rights] between the individual and the bank.

\subsubsection{Summary of all questionnaire results}
\label{sec:questionnaire_summary_final}

The results we gathered from the questionnaire recount that for the majority of the participants, the explanations provided in our case-study were not helpful to understand the decision nor suitable to assess its lawfulness.
In particular, we found that our explanations were generally considered difficult to understand, and incomplete and lacking in relevant and legally required information. 
{It was also generally considered that explanations did not} allow individuals to effectively exercise their rights. {T}hey were not found suitable or adequate {to allow} individuals affected by an automated decision to understand it, verify its lawfulness and fairness, and contest it if deemed appropriate. These perceptions towards the explanations we provided were made {regardless} of whether the decision was a true or false negative and the positive or negative impact they had on the bank's customers. 

It is also worth mentioning that no concerns were shown regarding possible conflicts of interests between the bank and its customers when providing the latter with XAI explanations about the automated decisions affecting them.

As clarified in Section~\ref{sec:follow-upinterviewssteps}, the script of the interviews was based on preliminary inferred conclusions reached upon the screening of the questionnaire results and the legal premises of our project. {The} outcomes we presented in this section were obtained after the follow-up interviews using a grounded-theory approach. Still, those preliminary inferred conclusions echo the assessment described {above}. In our preliminary conclusions we inferred the relevance of {the scope for an explanation to be understood, its legal compliance and the} potential lack of a conflict of interest between the parties involved in the automated decision-making {with regard to}~explainability. Hence, we developed a script through which the participant could offer further details, arguments, and discussion over {two problems}: 1) the challenge to understand the explanations shown due to the amount (or lack of) information provided, and the lack of helpfulness of such an explanation for individuals to understand the lawfulness of the decision and contest it if deemed appropriate, and 2) the possible lack of concerns regarding the interest of the data subject and the interest of the bank. 
{Therefore, while} not fully aligned with the conclusions reached after the in-depth analysis of the questionnaire responses, the inferred assumptions we used to script the interviews were partially unpolished and insufficiently detailed, {rather than discrepant.} They were obtained from a superficial evaluation so incompleteness was expected and assumed.

\subsection{Interviews}

\begin{figure}
    \centering
    \includegraphics[width=1\textwidth]{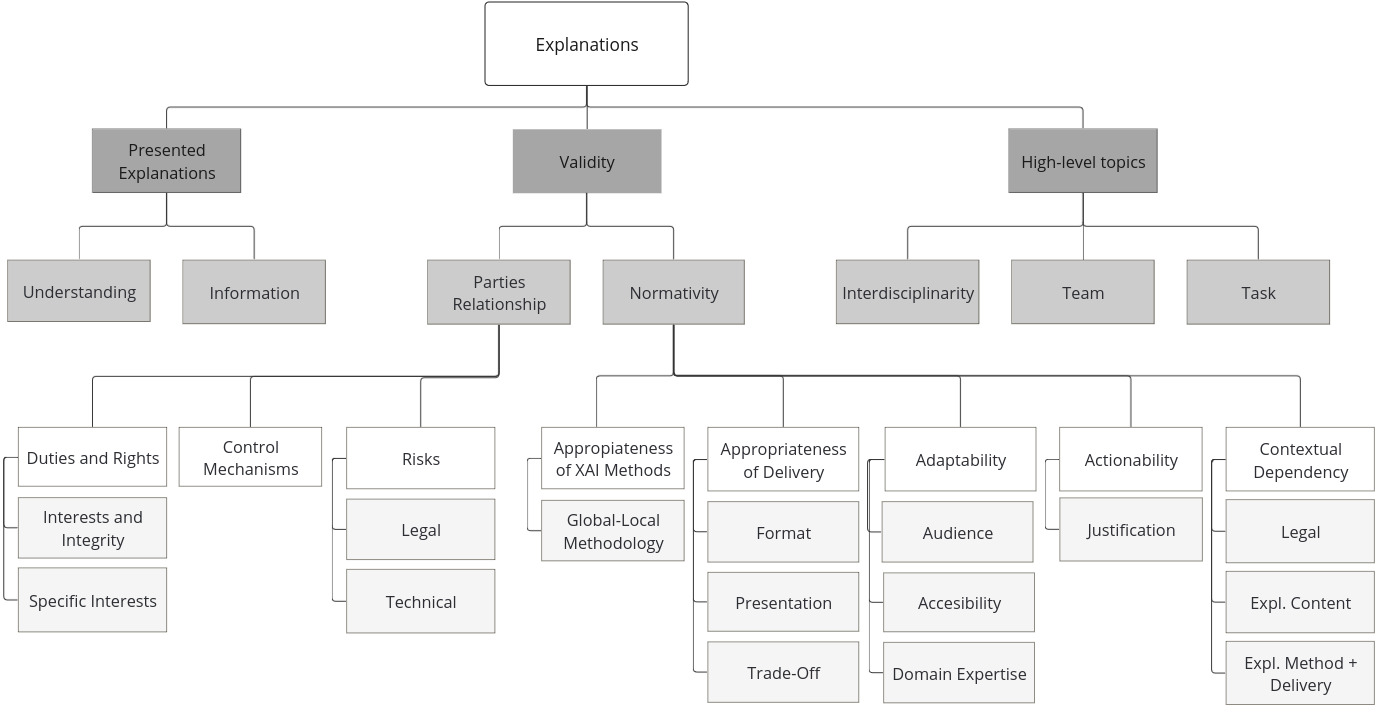}
    \caption{Hierarchy of codes. Dark gray boxes refer to core phenomena, gray boxes to themes, white boxes to sub-themes and light gray boxes to categories (top to bottom).}
    \label{fig:hierarchy_of_codes}
\end{figure}

\subsubsection{Summary of the coding table}
\label{sec:interview_summary_of_coding_table}

Here, we give an overview of the \textit{Aggregation Coding Table} that we developed from our interviews and via a grounded-theory approach. 
We only dive into definitions when necessary for the {reader's} general understanding. 
The table and all definitions can be found in the Appendix~\ref{sec:appendix_coding_tables}. 
We visualized the hierarchy arising from the coding table in Figure~\ref{fig:hierarchy_of_codes}.
From our analysis, we could identify three separate core phenomena:  
\begin{enumerate}
    \item \textit{Presented explanations}: the explanations we presented in the interview, and previously in the questionnaire; 
    \item \textit{Validity}: the standard to evaluate (the presented) explanations against, according to the applicable techno-legal framework; and
    \item \textit{High-level topics}: sub-themes that do not fall under the previous two but that emerged dynamically during the interviews and refer to high-level considerations about XAI. 
\end{enumerate}

\paragraph*{Presented explanations} The two themes identified are -- not surprisingly -- close to the preliminary conclusions from the questionnaire that we used to prepare the interview: a) \textit{understanding}, i.e.,~whether the explanation is easy or hard to understand; and b) \textit{information}, i.e.,~whether the information presented by the explanation is complete {or~not}.

\paragraph*{Validity}
This core phenomena is clustered into two themes: a) \textit{parties' relationship}, i.e.,~the relationship (here) between the data controller and subject
and b) \textit{normativity}, i.e.,~the way an explanation should be provided according to Article 22 of the GDPR. 

We divided these themes into several sub-themes.  
Parties' relationship (a) can be understood under the sub-themes of i) \textit{duties and rights}, with interests and integrity (axis: either aligned or conflicted for both), and specific interests (happy data subject, business) as the two categories. 
We further identified ii) \textit{risks}, and {draw} a distinction between legal and technical risks, as well as iii) \textit{control mechanisms}, with the category of oversight.

Normativity (b) is divided into i) \textit{appropriateness of the XAI method}, with the category of global-local methodology (axis: complimentary or exclusive).
A second identified sub-theme is ii) \textit{appropriateness of delivery}, with categories format (axis: complimentary or exclusive), presentation (axis: layered or static) and trade-off.
We further identified iii) \textit{adaptability}, with the category of audience, accessibility and domain expertise (axis: complete or missing).
We find iv) \textit{actionability} with the category of justification, and v) \textit{contextual dependency}, with categories of legal, explanation content, explanation method, and delivery.

\paragraph*{High-level topics} The core phenomena is divided into a) \textit{interdisciplinarity}, i.e.,~
approaching explanations from multiple perspectives; b) \textit{team}, i.e.,~{the design and implementation of} an explanation
with the participation of multiple stakeholders including the data subject; and c) \textit{task}, i.e.,~high-level discussions on the function of an explanation.

\subsubsection{Summary of the results}
\label{sec:interview_summary_of_results}

We follow the same sub-section~structure as above. {Additionally, we split the phenomenon \textit{validity} by its identified themes, to make it easier for the reader to follow.}
We {have added} some \textit{exemplary quotes} to support the analysis.

\paragraph{Presented explanations}

From the questionnaire, we found that explanations were in general hard to understand and that they {lacked} some information (see Section~\ref{sec:questionnaire_summary_final}).
We can confirm this finding via the interviews.
We identified answers to this core phenomena from three interview participants. Most of the answers refer to understanding while we identified one under the theme of information.
The explanations were found to be hard to understand (\enquote{it was very unclear what actually was going on in the pictures}), sometimes with reference to the \enquote{average person}, or to themselves (\enquote{but I am a lawyer}, \enquote{even me, who has read a paper or two}).
One participant also pointed out that the presented explanations are not yet final (\enquote{They are a very good intermediate level before the final product}).

\paragraph{Validity: parties' relationship}

\begin{itemize}
\item Under the sub-theme of \textbf{duties and rights}, which was further split into interests and integrity, and specific interests, we could find answers from all interview participants, suggesting and emphasizing different perspectives.
While some participants could not identify conflicts of interests (\enquote{I didn't perceive any risk for the competitive interest of the data controller}), others did (\enquote{conflicts in many dimensions}), with a slight tendency towards an agreement to no conflict.
This is in partial agreement with the results from the questionnaire (see Section~\ref{sec:questionnaire_summary_final}), in which we could identify no concerns about conflict of interests.
In identifying the sub-category of \textit{interests}, we could also find some statements about the relation between \textit{interests and legal rights and duties} (\enquote{do not want to reveal more than necessary}).
Participants pointed towards the fact that the provision of an explanation is also an act of balancing interests (\enquote{They should balance the interests of the bank with the ones of the clients}); with some emphasizing the interests of the data subjects, others {highlighted those of} the business. 
Therefore, a specific sub-category that repeatedly emerged was that of the \textit{happy data subject} (\enquote{protecting the data subjects}, \enquote{put yourself in the shoes of the data subject}, \enquote{consumer happiness}),
{in contrast} to the sub-category of \textit{business} (\enquote{played fair and exploit the fact of being fair}, \enquote{sell it so you can exploit}, \enquote{burden in an opportunity}, \enquote{secrecy demands}).
{Moreover}, participants pointed out how explanations may help to avoid a conflict of interest (\enquote{the well-framed simple explanation can resolve any potential conflicts of interests rather than amplify}, \enquote{anticipate legal issues from the design stage}).
Furthermore, one participant pointed to the hierarchy that exists between the data controller and subject (\enquote{data subject is the weaker party in this relationship}).
Others made some general comments about the concern of intellectual property (\enquote{the argument of intellectual property is very instrumental}, \enquote{but there are tools to protect it in case of infringement}).

\item The sub-theme of \textbf{risks} is divided into \textit{legal} (\enquote{This decision is at least based on sort of sensible things}) and \textit{technical} (\enquote{reverse engineering}) risks. While the first is closely tied to the above concept of duties and rights, the second establishes an additional connection to the technical details of an explanation.

\item The sub-theme of \textbf{control mechanisms} with the specific category of \textit{oversight} is derived from one interview (\enquote{independent and let's say a closed public or private body}, \enquote{a software tool}). 
\end{itemize}

\paragraph{Validity: normativity}

\begin{itemize}
    \item The sub-theme of \textbf{appropriateness of the XAI method} {encompasses} the category of global-local-methodology.
    Regarding preferences, we identify interview responses from two participants, both mentioning a combination (\enquote{complementary}, \enquote{global before and local afterwards}).
    This is somehow different from the result identified in the questionnaire in which we found both opinions {which entailed a} combination of global and local methods, as well as a preference towards local (see analysis of the comparison questions, Section~\ref{sec:questionnaire_comparison_questions_results}).
    \item The concept of \textbf{appropriateness of delivery} {encompasses} several categories. 
    Regarding the format of an explanation, we find answers from all interview participants. While the details diverge, we can identify an overall preference towards a combination of text with plots   (\enquote{I would combine the text with the image}, versus \enquote{more a narrative} and \enquote{visualize rather than reads}).
    We can further identify some concrete recommendations regarding the style of text and plots  (\enquote{very plain language}, \enquote{translate them into easier language}, \enquote{with the shape and colors}, \enquote{highlighting with like bold}).
    We also find some specific recommendations that may connect to explanation types such as contrastive explanations (\enquote{what would affect the decision making so that the decision would be different}, \enquote{strict conversation}),
    and a comment about how cognitive biases may affect the understanding of the explanation (\enquote{biases in the in reading numbers}). 
    The category of presentation is closely connected to that of format. We specifically use it to identify whether an explanation may be presented in a layered or static way. The first was prominently argued for by one interview participant (\enquote{layered explanation}, \enquote{for more info, click here}, \enquote{step by step}), while the other can be understood as \enquote{default}.
    Also, one interview participant explicitly mentioned the trade-off between accuracy and understandability one faces when presenting explanations (\enquote{trade-off between accuracy of explanation and understandability}, category trade-off).
    \item Under the sub-theme of \textbf{adaptability}, 
    we first discuss the audience (end-user, here the data subject and customer of the bank) of the explanations. Answers to this category can be derived from four interviews.
    While some general statements are made (\enquote{diverse pool of stakeholders with different expertises}, \enquote{depends on who is the recipient of the explanation}), we also find that participants identify different stakeholders and needs (\enquote{two different benchmarks [...] average person [...] average lawyer}, \enquote{someone developing this tool}).
    One participant also pointed out that the concept of an \enquote{average user} may be flawed (\enquote{I don't think that there is like an average}).
    Under the category of accessibility, we collect answers that go beyond that of the format and presentation of an explanation, but are close to the literal meaning of \enquote{accessibility} 
    (\enquote{could be easily [...] collected [...] by the clients}, \enquote{possible translation tools, hearings, impairments}, \enquote{consumer fatigue}).
    We also identified the category of (lacking) domain-knowledge (\enquote{domain expertise}, \enquote{implicit knowledge}).
    \item We introduce the sub-theme of \textbf{justification} because we found that in two interviews, participants prominently argue in favor of adding a justification (\enquote{Why?}) to the explanation (\enquote{How?}). 
    This also confirms findings from the {questionnaire (see Section~\ref{sec:questionnaire_summary_final}).}
    Next to mentioning \enquote{justification} explicitly, keywords that emerge here are law/lawfulness (\enquote{that decision making is lawful and respect some laws}), as well as 
    contestation and argumentation (\enquote{the lines of argumentation that you would want to make as a person are not the lines of argumentation that are in most of these.}, \enquote{make decision contestable}).
    \item
    The last sub-theme {identified} is \textbf{contextual dependency}. This sub-theme has to be {understood broadly}, and we identify answers from four interview participants. 
    Many different statements could be clustered into this sub-theme (\enquote{Facts, norms, and [...] the context of the decision}, \enquote{on the type of the decision, the [...] type or immediacy of the remedies.}, \enquote{more significant the [...] decision is to a person, I think the more level of detail}, \enquote{depends on when now we apply them}, \enquote{relevant legal framework},\enquote{very sectorial}, \enquote{intuitive guess that the [...] local one is is better, but depending on the context}).
    From these statements, we may broadly infer a dependency \textit{of} (what is dependent) 
    legal norms and rules;
    the explanation content;the explanation method and delivery (the three topics extracted),
    and a dependency \textit{on} (dependent on what)
    application context including the sector and country;
    significance of the decision; and
    (again) legal norms and rules.
    We also identified once {the lack of such} context (\enquote{lacking the context}).
\end{itemize}

\paragraph{High-level topics}

We identified three themes: 
interdisciplinarity (\enquote{goes out of the field of explainable AI}, \enquote{multidisciplinary}, \enquote{Experts working together.}),
team (\enquote{start talking to consumers}, \enquote{not to leave [...] the design choices to complete the engineers}, \enquote{in the feet of, like, lawyers}),
as well as task (\enquote{explanation is not meant for lawfulness. Understanding it's usually meant for causality understanding}, \enquote{As a lawyer, I our main concern would be is there a efficient process procedure irrespective of the understandability of the rules whatsoever}, \enquote{contestability {[...]}
do not help with it very much
}).

\subsection{Relevant interconnections}

In this section we reflect on two connections between the different codes derived and explained above, which 
set the basic grounds for the conclusions presented in Section~\ref{sec:discussion}.

\paragraph*{Context of an explanation}
Explanations depend on the context and the targeted end-user \citep{gunning2019xai,DBLP:conf/fat/MittelstadtRW19}.
Based on the interviews, 
we could identify the category of \textit{audience}
and the sub-theme of \textit{contextual dependency}.
These two codes are strictly connected to each other.
Specifically, we understand \textit{contextual dependency} as a much broader topic than \textit{audience}, with the first encompassing the sector, country and region, the legal framework and its specific application {in the context in which} the automated decision-making takes place, 
and the latter referring to the adaptation of the automated decision's explanation {in respect} of its particular audience, i.e.,~recipient of the explanation.

We connect both topics by remitting to Section~\ref{sec:interview_summary_of_coding_table}: explanations depend on a series of elements particular to the automated decision-making process, such as the sector and country {in which} it is implemented, the significance and outcome of the decision, and the particular norms and rules that apply to that concrete situation. 
However, the explanations are also dependent on other aspects that introduce 
limitations and restriction, such as the 
methods and formats of delivery, their 
content, or the requirements established via legal norms and rules. The category of audience is part of the former aspects influencing the explanations.

\paragraph*{Trade-offs}
{Connecting the codes derived from the}
interviews, we 
{identified} two central trade-offs:
1) the trade-off between the amount of information presented in an explanation and its understandability; and 2) the trade-off between the interests and rights of the data controller (the bank), and the interests and rights of the data subject (the client).

The first trade-off is well-known and discussed in the (technical) literature on XAI \citep{DBLP:journals/csur/NautaTPNPSSKS23}. A possible remedy to it -- arising from our study -- may be the use of a layered approach to present an explanation, or a presentation via interaction tailored to the necessities and desires of the {end-user.}

The second trade-off is similarly well known. As already discussed in Section~\ref{sec:interview_summary_of_results},
different opinions on which party should be in focus, exist. 
One remedy for this trade-off may be to assert its control -- we also identified the topic of \textit{control mechanisms} through our expert study. 
Further, we found that participants
either suggested that an explanation itself may be (a part of the) solution for this conflict,
or that an explanation could potentially consolidate the conflict, tilting the balance negatively towards the data subject.
It further connects the discussion to the identified code of \textit{task}, under which we summarized the discussion of the task and function of an explanation.
The two nuances mirror parts of the current debate about the deployment of XAI methods, with a focus on regulation.
We mentioned the legal debate on the \enquote{right to an explanation} in the GDPR already in Section~\ref{sec:legal_background}.

\section{Discussion}
\label{sec:discussion}

\subsection{Answering the research questions}
\label{sec:conclusion_rq}

In this section, we answer {the} research questions {of this study} (see Section~\ref{sec:research_questions}).
{We}
point to the relevant sections for in-detail information.

\paragraph{RQ1}

The explanations we presented to the legal experts were not complete in the information they displayed (see both Section~\ref{sec:questionnaire_summary_final}, and Section~\ref{sec:interview_summary_of_results}).
Global SHAP explanations
were deemed {insufficient} as they 
focused on the global description of the decision-making process rather than the particular case of the individual affected. 
{For completeness,} Global SHAP would need to be accompanied by a local explanation providing information pertinent to the individual and the specifics of the actual case. However, the local XAI methods used in our scenario were {also deemed to be}
incomplete to {a certain} degree. DICE explanations were
appreciated as they seemed to clarify the features deciding the particular case, but they were {also} treated with caution.
It was not easy to understand how {DICE} works nor {did the method} seem to offer actionable information beyond the possible change of behavior or adjustment on personal conditions. 
Local SHAP explanations might not be the best addition to its global counterpart as they were perceived {as being} too complex in their format (graphical) and missing information regarding the single features determining the decision, which was {also} among the missing information in Global SHAP. LORE explanations, although receiving a positive response, coincided in the same weakness as the
{other local}
XAI methods{; that is to say,} they were unclear in their format and lacked information about the decisive features' benchmark, and the effect a change in circumstances {has on} the decision.

Further, XAI methods used in our scenario
{provided neither enough nor} suitable information in terms of the individuals' right to actionability. 
{This was due}
the lack of particular information about the concrete case, the complexity in their format and delivery method and its subsequent limited understanding for an average individual, or because they were not strictly explanations as far as the intent of the GDPR is concerned. 
{This last point holds quite heavily for contrastive explanations:}
a controversial point was brought up in the questionnaire regarding whether contrastive explanations about what features would need to change in order to get the opposite decision shall or shall not be considered as suitable information to contest a decision or assess its lawfulness and fairness. 
We 
{found}
that the compliance of XAI explanations to the GDPR strongly depends on {how they allow individuals to exercise their rights}

We can deduce that a combination of global and local XAI methods was preferable over the use of a single method; {such a combination} would provide a more complete and actionable explanation about the automated decision. 
{Furthermore, }besides the information 
provided by the XAI methods, a text or narrative was deemed highly necessary, particularly explaining the XAI plots  and tables and addressing the motives and reasons behind the decision. We could not identify a specific part of the explanation as being more relevant than another.
Finally, it can be stated that the reasoning did not significantly change between Case TP and Case FP, i.e., the correctness of the outcome did not have any special impact on the {responses given}.

\paragraph{RQ2}

The explanations were hard to understand by the experts, and (therefore) their understanding of the explanations was limited as can be seen both in Section~\ref{sec:questionnaire_summary_final}, and Section~\ref{sec:interview_summary_of_results}.

Some of the XAI explanations 
were described as possibly cognitively misleading and confusing, demonstrating a certain level of untruthfulness. 
At the same time, however, such XAI explanations were seen {to provide} an opportunity for the data controller to play fair and exploit their information duties, and center the needs of the data subject.
Furthermore, correctly designed and intended explanations could prevent any potential conflict of interest between controllers and subjects, anticipating legal disagreements and complaints. 
In that regard, although our XAI explanations 
{were met with} some hesitation and skepticism, better constructed explanations -- 
as discussed in Section~\ref{sec:questionnaire_summary_final} and Section~\ref{sec:developers} -- may be 
received with more confidence and trust. 

\paragraph*{High-level topics}
While not strictly answering our research questions, we also summarize results connected to the identified high-level topics
(\textit{interdisciplinarity}, \textit{team} and \textit{task}).
We have discussed the topic of \textit{task} under the trade-off between the interests and rights of the data subject and the controller. Delineating the \textit{task} of an explanation is crucial to construct an explanation in the first place and to understand its potential and limits. 
We add that beyond a \textit{task} in the relationship between the data subject and controller in this scenario, an explanation can have tasks that, for example, aim for an understanding for {the purposes of facilitating} progress in science and industry, or for debugging purposes \citep{DBLP:conf/iclp/State21}.
The topic of \textit{team} will be discussed below and in conjunction with a recommendation about {adapting} explanations to end-users, and incorporating them into the design process.
As for \textit{interdisciplinarity}, we re-emphasize the need for different expertise and perspectives in the field of XAI.

\subsection{Recommendations to improve existing explanations}

In this section, we present the implications from our expert focus study. We extract some concrete recommendations for developers in Section~\ref{sec:developers}, and point to relevant legal issues and open debates in Section~\ref{sec:legal_pointers}.
These two sections are firmly grounded on both the questionnaire and interview. 
Therefore, it is important to remember that our expert study is tied (only) to model-agnostic explanation methods, and to a specific loan application scenario that targets a lay end-user as the receiver of the explanations. Likewise, the poll of experts who participated in our project was small, and results should thus be regarded with caution and as a limited illustration of how legal experts reason about explanations for AI systems, and how they judge their legal compliance. That {being} considered, we can draw some interesting thoughts. 
We add relevant literature and references to connected topics, if applicable.

\subsubsection{Developing and designing the explanations}
\label{sec:developers}

Here, we discuss specific recommendations that we drew from both the questionnaire and the interview (specifically, Section~\ref{sec:interview_summary_of_results}, presented explanations and validity). 
These recommendations are mainly targeted at developers of explanations.

\paragraph{Presentation and format of explanations, and the choice of XAI method}

{To improve understandability}, explanations may feature a combination of global and local methods (and information), and a combination of plots  and texts.
Texts should draw on easy language, math should in general be avoided, colors and bold markers in the plots  and texts may be used to support the presentation of the explanation.
Moreover, a layered structure of the explanation may be suitable, {as might a degree of} interactivity and a conversational style of the explanation. {Further,} the use of contrastive explanation methods
{-- beyond the two methods discussed in this study, presented in Section~\ref{sec:explainable_ai} and shown in the Appendix~\ref{sec:appendix_explanations} -- may be suitable}.
\vspace{0.1cm} \\
We refer to the following \textbf{literature}: interactivity and explanations~\citep{DBLP:journals/ai/Miller19,DBLP:journals/corr/abs-1712-00547,DBLP:journals/cacm/WeldB19}, combining different XAI methods~\citep{LONGO2024102301}, the introduction of contrastive explanations in XAI~\citep{wachter2018a}, and surveys on contrastive explanations~\citep{DBLP:journals/corr/abs-2010-04050, Guidotti2022, DBLP:conf/ijcai/KeaneKDS21, DBLP:journals/access/StepinACP21}.

\paragraph{Accessibility of explanations}
Explanations may provide translations of texts into several languages, as well as measures to make them accessible for visually impaired people, and more inclusive in general. 
A {step towards better accessibility} could {encompass} an additional provision of an audio {tool} that describes the textual or graphical explanation.
Further, the end-user of the explanation must be aware of where to find the explanation in the first place.

\paragraph{Content of an explanation} Explanations may profit from additional domain or background knowledge, and from adding the reasons and motives behind the decision.
\vspace{0.1cm}\\
We refer to the following \textbf{literature}: background knowledge and explanations~\citep{DBLP:journals/corr/abs-2105-10172,DBLP:conf/iclp/State21}, as well as a discussion of the concept of \textit{justification}~\citep{malgieri2021just,malgieri2017right}.

\paragraph{Technical risks of explanations}
The main identified technical risk was reverse engineering. 
We point to~\citep{DBLP:conf/fat/PaniguttiHHLYJS23,DBLP:conf/fat/NanniniBS23} providing not only a discussion of the role of explanations in regulation but also of their technical shortcomings, beyond those mentioned in this paper, and~\citep{DBLP:conf/fat/BordtFRL22}, a position paper on post-hoc explainability methods. 
Technical risks should not necessarily be taken as a sign to abstain from the use of explanations but to use them with care, both in relation to what an explanation can do for the end-user (e.g., providing faithful, reliable information versus skewed, unreliable information in relation to the risk of technical faithfulness and stability) and to the provider (e.g., providing information via a pipeline that may be easy or hard to jeopardize, in relation to the technical risk of reverse engineering). Further, some technical risks may be resolved in the future by research efforts, or (partly) counteracted by regulatory efforts itself.
Therefore, regulatory efforts must be carefully aware of limits of XAI methods.

\paragraph{End-users of explanations and their needs}
User studies are necessary to assess state-of-the-art explanations
and to understand better the individual needs of the targeted end-users. We refer to Section~\ref{sec:user_studies} for more information.
While we did not specifically discuss user studies in this work, we can draw a close connection to the category of \textit{audience}, emphasizing the perspective and the needs of the end-user of an explanation, and arguing for their integration.
A categorization of different audiences of XAI that reflects this idea, is presented in~\citep{suresh2021beyond}.
A second connection can be made to the theme of \textit{team}, pointing towards participatory design approaches~\citep{asaro2014participatory,book}.

\subsubsection{Addressing legal uncertainties, loopholes, and questions}
\label{sec:legal_pointers}

Here, we address some of the legal premises we assumed during the design of our case-scenario and discuss open questions on the practical compliance {to the right}
to contest and to obtain an explanation about an automated decision (see Section~\ref{sec:legal_background}). 
We draw this section 
on the loopholes, uncertainties and questions regarding Article 22 of the GDPR
from both the questionnaire and the interview, and
{primarily target}
practitioners and legal experts{; it is submitted that} developers can also benefit from these considerations.

\paragraph{Understandability and interpretability to ensure contestability}
Article 22 of the GDPR entitles data subjects to contest {an} automated decision {that} legally or similarly affect{s} them.
Consequently, individuals shall know the motives and reasons behind such a decision. The precondition to contest an automated decision is to understand \textit{how} the model works and \textit{why} that particular decision was reached~\citep{malgieri2021just, malgieri2017right}. If the explanation does not clearly answer these questions, it would likely be deemed \textit{inappropriate} as
{it is} {not} suitable to {allow individuals to} exercise 
{their} rights.

\paragraph{The (lack of) conditionality of the transparency thresholds}
The threshold of information provided about an automated decision is independent of the outcome of that decision and the positive or negative effects it might have on the individual. Individuals might be less demanding when a positive decision affects them 
{as} they could {find} the exercise of their 
rights to be redundant. However, they may still {desire} to verify the lawfulness, fairness, and accuracy of such decisions for {which} clear, complete and understandable explanations are necessary.

\paragraph{Intellectual and industry rights and secrecy {versus rights of the individual}, the two sides of a coin}
Corporate secrecy, and intellectual and industry rights can raise problems with regard to {the transparency of} automated decision-making, as they may impede and obstruct the possibility of providing information about the system and its decisions. 
Secrecy -- understood as an umbrella term -- can limit the amount of information the data controller is compelled to provide, but such limitations {would generally be deemed to be justified}. The possible burden of providing information could be, nonetheless, beneficial for data controllers if they show their predisposition to stretch their own secrecy limits, for example, {by} putting in place ethical benchmarks or transparency codes. In any case, the right to contest and to obtain an explanation shall not be unbearably undermined in the name of secrecy{; this would be deemed to be} unjustifiable. 

\paragraph{Format and presentation of {provided} information}
Information about automated decisions shall be concise, transparent, intelligible and easily accessible as referred to in Article 12(1) of the GDPR. However, XAI methods and techniques provide -- in general -- quite technical and complex information about the system and the decisions reached. 
Hence, XAI explanations {can} be difficult to understand even for a legal expert, so for an average individual, their understanding could be even more challenging. We thus remit to Section~\ref{sec:developers} for practical guidance and recommendations, although we need to express some caution as 
{undoubtedly}, there is no unique combination of presentation, format and method that suits all the cases where information is required. Trying to do so could result in a repetition of the \textit{cookie pop-ups problem} where cookie consent mechanisms ended up in nudging and influencing people's privacy choice, despite their original purpose of empowering them~\citep{kretschmer2021cookie,machuletz2019multiple,matte2020cookie}. 

\paragraph{XAI methods to provide meaningful information about the automated decision and its significance and envisaged consequences} 

As presented in Section~\ref{sec:legal_background}, the Article 29 Working Party offered some recommendations on how to comply with the transparency provisions of the GDPR. {However,} numerous uncertainties remain about how to effectively comply with {the information and explanation requirements of the GDPR}. 
The whole purpose of this {study} is to offer some insights in this regard, but here we will refer to Sections~\ref{sec:conclusion_rq} and \ref{sec:developers} {to} suggest {different and possible options for the} practical materialization of the recommendations of the Article 29 Working Party.

\paragraph{Parties' relationships set expectations and legal grounds} 
An automated decision takes place in a context {that is} delimited {by the} relationship between the parties, either arisen from the direct consent, the necessity of a contract, or the authorization of a EU Member State. 
This relationship already delimits some of the rights and duties that both parties shall respect and may enjoy. Explanations about automated decisions are not isolated legal islands. Article 22 of the GDPR sets out the basis for the provision of information, but other applicable norms may need to be taken into {account} when explaining automated decisions.  

\section{Limitations and future work}
\label{sec:limiations}

This study has several limitations. 
A well-known limit of the grounded theory approach is the interpretive nature of coding, which introduces the potential for researcher subjectivity bias impacting the objectivity of the generated theories. 
To overcome this issue, two researchers with different perspectives, namely, legal and technical, performed the coding process with the aim of achieving a consensus, and built on strong documentation efforts throughout their analysis. 
However, to support the validity of our findings, future experimental research is needed.    
{Furthermore, }we used purposive sampling to reach out to our target group of legal experts. This sampling strategy is determined by our professional network, and therefore {introduces some biases}.
{Thus, most of the participating legal experts were academics. However, while from the interviews, we have an idea about the background of our participants, we do not know the background of the participants in the questionnaire as we did not ask about it.}
Also, the number of participants {in our study was small}.
As discussed in Section~\ref{sec:participants_selection_validity}, however, this expert focus study does not claim {to provide} for general validity but rather {to allow for the identification of issues surrounding} the legal validity of explanations.
{Furthermore, we acknowledge an additional bias that potentially affects the validity of our results by our framing of the questions, specifically by using leading questions, and by the structure of the multiple choice questions - a Likert scale with five options.}

Beyond experimental limitations, we point to the following:
the loan application scenario constructed for this {study} assumed that the bank, {giving out} explanations to its customers, is providing them truthfully. We point to this in the questionnaire.
We understand that this assumption is simplistic and overlooks possible risks such as an intentional misguiding of the customer via the explanations to the disadvantage of the customer and the advantage of the bank. 
However, \textit{firstly} we drew on this simplistic assumption for research purposes only, being well aware of its shortcomings.
\textit{Secondly} we avoided presenting a scenario where the bank acts in bad faith or misconduct as it would
create {a reluctance} for the legal experts towards the explanations' actual lawfulness and compliance. Assuming the truthfulness and good faith of the bank allows for the evaluation of explanations {with regard to}~their limitations and challenges without {highlighting unduly how} possible malpractice could affect the reasoning and assessments of the experts. We acknowledge the limitation this assumption poses, thus, encouraging further research on how XAI methods could (not) be potentially used to trick explainability and information requirements. 

This work was situated in the European legal context with a specific focus on the GDPR. Only through this \enquote{situation}, we made it possible to derive the specific opinions of legal experts towards XAI as referred to in the pertinent provisions of the GDPR. However, we did not consider other potentially relevant legislation, such as the European AI Act or consumer protection law. Nor did we encourage our participants to further dwell on other legislative requirements on algorithmic explainability and transparency beyond what would be blatant or unavoidable {reference} to applicable laws and regulation \cite{DSA,DMA,aiact}.
Therefore, extending the study to a more encompassing case-study where different explainability and information requirements could overlap is a point of future research. 

{We entertain such a possibility with regard to the right to an explanation for individual decision-making as referred to in Article 86 of the AI Act. However, this paper could not focus on that provision given the uncertainty and uneasiness of basing the legal premise of our study on a right that had not yet been formally included in a European law.  We acknowledge the importance and applicability such research has as well as the synergies that can be found with our paper.}

Last, we would also like to point to the fact that this study was focusing on a specific type of explanation -- global and local, model-agnostic explanations -- and only on a selected subset of four explanations. As discussed above, this choice is to some extent based on the popularity of these methods. Moreover, it builds on our own (legal and technical) expertise. 
An extension of this study design to other types of XAI methods such as model-specific methods, and including a wider range of methods in general is therefore desirable and needed. Furthermore, based on our insights we find it quite relevant -- and therefore advocate for -- more extensive work on explanations that combine two or more types of XAI methods, particularly if they combine both global and local, or even model-agnostic and model-specific methods.

\section{Conclusion}
\label{sec:conclusion}

In this paper, we presented the \textit{explanation dialogues}, an expert focus study to understand the perception, expectation and reasoning of legal scholars towards technical explanations for ADM systems, and as provided by XAI methods.
Our study consisted of two parts: an online questionnaire, and a follow-up interview, both guided in their evaluation by grounded theory.

Our results are two-fold: \textit{first}, we presented both a hierarchical and interconnected set of codes, {whereby} each {code was associated} with a specific definition that arose from the study, and a summary of the standpoints of the participants of our study.
While we found that the presented state-of-the-art XAI methods face both shortcomings in terms of their understandability, presented information, and suitability to exercise the rights {with regard to} the data subject and the controller, we also discussed issues that may arise from the possibly different interests of the data controller and subject.
Regarding the legal compliance of the presented explanations, the result is mixed, i.e., we cannot state that one method we presented complies with the GDPR. 
{However, we can assert that the perceived conformity of the explanations with the GDPR is closely connected to how they allow individuals to exercise their rights.}
We further found that while the interviewed legal experts are well informed {as regards} explainability, unsurprisingly, they may have some knowledge gaps regarding technical properties {of explanations}.
\textit{Second,}
we identified specific recommendations for technical practitioners, and {indications of} legal areas of discussion, both pointing towards avenues for future research.

While the insights we gathered through this study show clear connections to existing state-of-the-art literature and debates on explainability, our study uniquely offers insights into the reasoning of legal experts presented with state-of-the-art XAI methods -- a study design that is, to the best of our knowledge, unique.
In summary, we identified different positions towards XAI methods, ranging from skeptical stances towards optimism, mirroring the current debate in this research area.

The work has a highly interdisciplinary character, situated between law, computer science, and social science.  
Furthermore, by situating the study in the EU legislative context and drawing on a specific use-case in the credit domain, we 
acknowledge the \textit{contextual} dependencies of work in the field of XAI. 

\paragraph*{Acknowledgments}
This work was supported by the European Union's Horizon 2020 research and innovation programme under Marie Sklodowska-Curie Actions for the project NoBIAS (g.a. No. 860630) and under the Excellent Science European Research Council (ERC) programme for the XAI project (g.a. No. 834756). Furthermore, the work of L. State, A. Beretta and S. Ruggieri has been partly funded by PNRR - M4C2 - Investimento 1.3, Partenariato Esteso PE00000013 - \enquote{FAIR - Future Artificial Intelligence Research} - Spoke 1 \enquote{Human-centered AI}, funded by the European Commission under the NextGeneration EU programme. This work reflects only the authors' views and the European Research Executive Agency (REA) is not responsible for any use that may be made of the information it contains.

\paragraph*{Author Contributions}
\textit{Laura State} conceptualization of study, questionnaire (design, data evaluation), interviews (script design, implementation, data evaluation), paper (first draft (major), revision), contact to participants, project organization/lead. \textit{Alejandra Bringas Colmenarejo} conceptualization of study, ethics approval, questionnaire (improvement, data evaluation), interviews (improvement script, implementation, data evaluation), paper (first draft (major), revision), contact to participants. \textit{Andrea Beretta} ethics approval (support), questionnaire (improvement, implementation, data cleaning), interview (improvement script), paper (first draft (minor), revision). \textit{Salvatore Ruggieri} conceptualization of study, questionnaire (improvement), interview (improvement script), paper revision, contact to participants. \textit{Franco Turini} conceptualization of study, paper revision. \textit{Stephanie Law} conceptualization of study, paper revision, contact to participants.

\paragraph*{Data Availability}
The questionnaire is anonymous, the interview data was anonymized, both are stored safely for 5 years.

\paragraph*{Code Availability}
We use standard code to provide the explanations, and evaluate the data. This code is therefore not public.

\paragraph*{Declarations}

\textbf{Conflict of interest}
The authors have no relevant financial or non-financial interests to disclose.

\noindent\textbf{Ethical approval}
The study was conducted in accordance with the Declaration of Helsinki and approved by the Faculty Ethics Committee of the University of Southampton, 27 March 2023 - ERGO number 80482.

\noindent\textbf{Consent to participate and to publish}
Informed consent was obtained from all individual participants included in the study. Participants were informed about the use of the questionnaire/interview data for an article. We do not publish any identifying information of the participants.

\bibliography{library}

\newpage
\begin{appendices}
\renewcommand\theHtable{AABB\arabic{table}}
\renewcommand\theHfigure{AABB\arabic{figure}}

\section{Questionnaire and interview details}

\subsection{Research questions and connection to questionnaire and interviews}
\label{sec:appendix_research_questions_design}

In Table \ref{tab:rq_versus_desgin}, we summarize how each RQ is incorporated into the design of both the questionnaire and the interviews.

\begin{table}[]
    \centering
    \begin{tabular}{llll}
    \toprule
        RQ && Questionnaire & Interview (*) \\
    \midrule
        \multirow{3}{*}{\textbf{1}} & (a) & question 3-4 & \multirow{3}{*}{second block main part}\\
        & (b)  & question 5 + comparison questions & \\
        & (c)  & question 6 &\\
        \multirow{2}{*}{\textbf{2}} & (a) & question 1-2 & \multirow{2}{*}{first block main part, cool-off} \\
        & (b)  & all questions, focus on comparison questions &\\
        \bottomrule
    \end{tabular}
    \caption{Research questions versus questionnaire and interview. (*) For the interview, we focus on the research questions as a whole, thus there is no mapping onto the subquestions.}
    \label{tab:rq_versus_desgin}
\end{table}

\subsection{Explanations}
\label{sec:appendix_explanations}

Here, we present explanations from our expert study. {We add textual explanations, building on the explanatory statements that we also presented in the questionnaire.}

{
\paragraph*{SHAP}
We present SHAP for Case TP in Figure \ref{fig:shap_local}.
The \textbf{global} explanation (upper panel) depicts the average contribution of a feature towards the outcome. The higher the value the higher the contribution of a feature.
The color differentiates the contribution with respect to the predicted class.
\newline
The \textbf{local} explanation (middle and lower panel) contains two plots, as both possible outcomes (low credit risk score/good credit/credit accepted and high credit risk score/bad credit/credit declined) are explained separately. Red values indicate positive SHAP values, blue values indicate negative SHAP values.
The data instance was correctly predicted as not eligible for credit.
\newline
The middle plot shows the explanation for \enquote{good credit}. Thus, starting from an average ADM risk score value for \enquote{good credit} of $0.7$ (the base value in the plot), the contributions of the different features (local SHAP values) push the ADM towards the final output value of $0.32$.
The lower plot shows the explanation for \enquote{bad credit} (predicted outcome).
Starting from an average ADM risk score value for \enquote{bad credit} of $0.3$ (the base value in the plot), the contributions of the different features (local SHAP values) push the ADM towards the final output value of $0.68$. 
The higher output value determines the final decision of “bad credit” (i.e., $0.68 > 0.32$).
}

{
\paragraph*{DICE}
We present DICE for Case TP in Figure \ref{fig:dice}.
The table holds the information about the data instance explained (column 2) and two contrastive (counterfactual) instances (column 3 and 4), which represent the explanation as provided by DICE. Attribute names are depicted in column 1.
We always show the level of a attribute (the actual value of the attribute) and the associated category (the meaning of the level).
Attributes that changed are marked in color.
}

{
\paragraph*{LORE}
We present LORE for Case TP in Figure \ref{fig:lore}.
LORE explains the decision of the ADM via a factual and a contrastive (counterfactual) decision rule.
These rules are extended by a statement about how to interpret them.
}

\renewcommand{\thefigure}{5}
\begin{figure}
    \centering
    \includegraphics[width=0.6\textwidth]{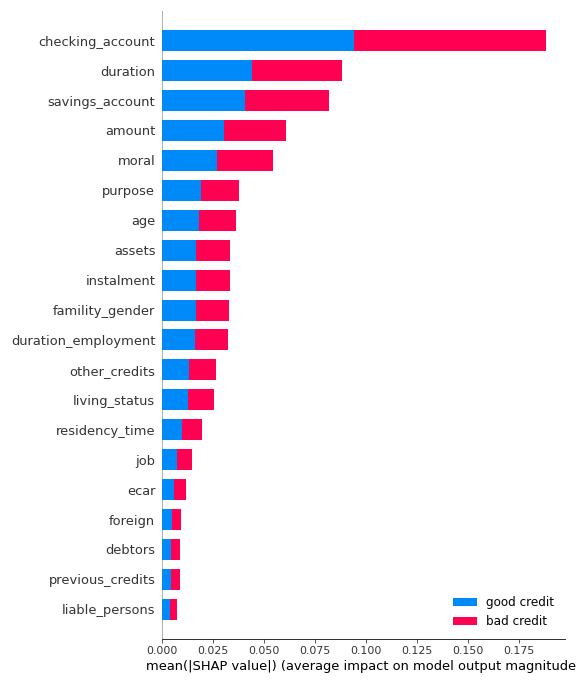}
    \includegraphics[width=.75\textwidth]{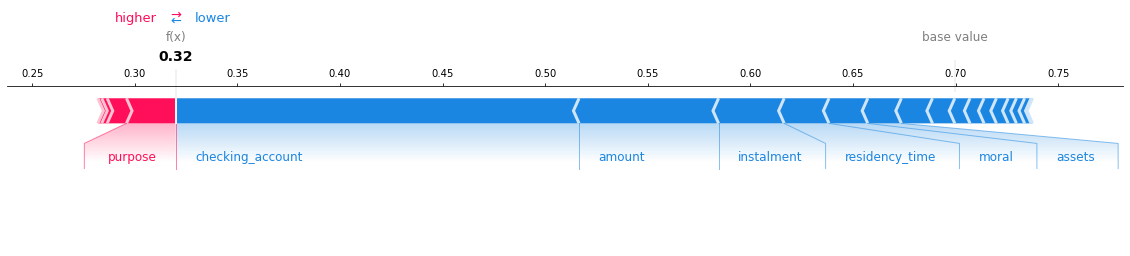}
    \includegraphics[width=.75\textwidth]{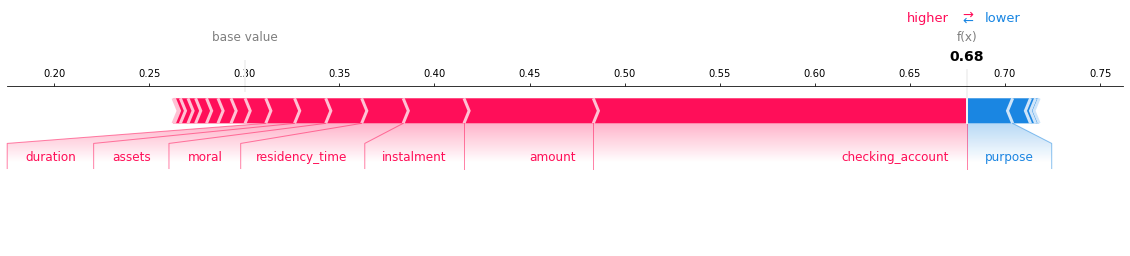}
    \caption{SHAP explanations. Upper: global SHAP, middle and lower: local SHAP for Case TP local (middle panel w.r.t. a low credit risk score/good credit, lower panel w.r.t. a high credit risk score/bad credit).}
    \label{fig:shap_local}
\end{figure}

\renewcommand{\thefigure}{6}
\begin{figure}
    \centering
    \includegraphics[width=.8\textwidth]{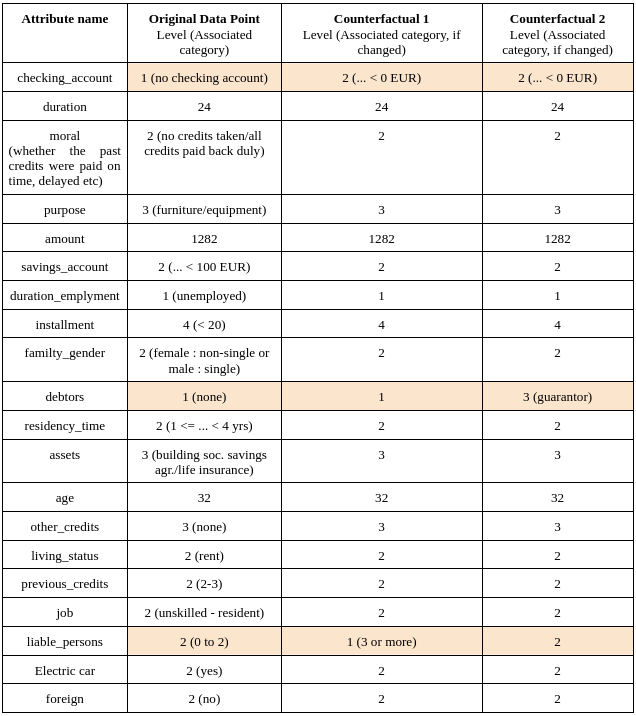}
    \caption{DICE explanation, presented via a table, for Case TP.}
    \label{fig:dice}
\end{figure}

\renewcommand{\thefigure}{7}
\begin{figure}
    \centering
    \includegraphics[width=.8\textwidth]{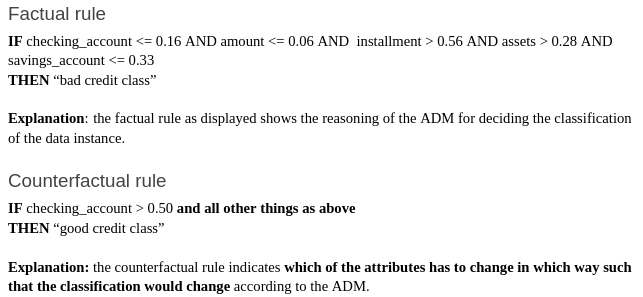}
    \caption{LORE explanation for Case TP.}
    \label{fig:lore}
\end{figure}

\subsection{Questionnaire}
\label{sec:appendix_questionnaire}

\subsubsection{Questions presented after a single explanation}

Here, we present the question that we gave to our participants after the presentation of a single explanation.

\textit{Intro text} We ask you to answer the following set of questions. Please remember your role: You are asked to answer the questions in the role of the internal consultant of the bank. We expect you to review the compliance of the information and explanations in regard to the interest and duties of the bank, and the interests and rights of the data subject.

\begin{enumerate}
\item How did this explanation contribute to your understanding of the ADM? (You can use the following questions as guidance: Do you understand the decision variables? Are they complete/informative/relevant? Was the format of the explanations helpful to you?)
\item Judge how well you understand the decision based on the shown explanation on the
following scale:
Not at all / A little bit / I have some questions / Good, but I might have missed something / Very well
\item  Is there any information you consider necessary to ensure your understanding of the
decision that is not included in the explanation presented? No / Yes (please detail your answer)
\item Is there any information in the explanation that you find confusing or misleading? No / Yes (please detail your answer)
\item Do you agree with the following statements? Strongly disagree / Disagree / Neither agree nor disagree / Agree / Strongly agree
\begin{itemize}
    \item The explanation presented was clear and understandable for an “average individual”.
    \item The explanation presented provides sufficient information about the decision to ensure
that the person affected by the decision understands the reasons and motives behind
it.
    \item The explanation presented allows the data subject to effectively exercise her right to
contest such automated decisions [if deemed appropriate], in accordance with the
safeguards established in Article 22 paragraphs 3 of the GDPR.
    \item The explanation provided does not pose a [potential] conflict with the interest and
rights of the bank (e.g., intellectual property).
    \item The explanation provided allows the data subject to verify the lawfulness and fairness
of the automated decision in regard to other sectoral laws applicable to the case (e.g.
consumer law, contract law, anti-discrimination law)
\end{itemize}
\item Regarding option a - e, please take a bit of time to explain your answers in a short text.
\item Would your assessment change if the outcome of the ADM system was different?
Please answer with a short text.
\end{enumerate}

\subsubsection{Questions presented at the end of a case (comparison questions)}

Here, we present the comparison questions.

\textit{Intro text} Before you move to the next case, we would like to ask you two more questions that
will help us to compare the previously shown explanations among each other.

\begin{enumerate}
\item Did you prefer one/several explanations over other explanations (for example, global,
local, or a specific method)? If yes, explain in a few words why.
\item If you’d be free to mix the explanations that were shown to you as you’d wish (type of
explanation/format and presentation of the explanation, etc.), how would you do that?
\end{enumerate}

\subsection{Dataset}
\label{sec:appendix_dataset}

Here, we discuss the dataset underlying to the questionnaire and the computed explanations.

We use the updated version of the South German Credit Dataset, which is a corrected version of this popular dataset.\footnote{\url{https://archive.ics.uci.edu/ml/datasets/South+German+Credit+(UPDATE)}, and \url{http://www1.beuth-hochschule.de/FB_II/reports/Report-2019-004.pdf}}

The dataset contains 1000 data rows, with each 20 features and a binary response variable (``good vs. bad'' credit) with a ratio of 7:3. The data was collected between 1973 and 1975, in Germany. Bad credits are oversampled from a base rate of $5 \%$
Nominal variables are associated to integer values, such that higher values correspond to a ``better'' credit level \citep{Groemping2019}.

We did the following data manipulations to account for the fact that the dataset is from the 70s: change of the currency from \enquote{DM} (Deutsche Mark, former German currency) to \enquote{EUR}, and change of the asset from \enquote{telephone} to understand the status of the applicant to \enquote{electric car} (electric cars as signifier of wealth/certain status).

\section{Coding tables}
\label{sec:appendix_coding_tables}

We present here the questionnaire coding table (Table~\ref{tab:questionnaire_coding_table_0}), as well as the aggregation coding table (split in three, Table \ref{tab:aggregation_coding_table_1}, \ref{tab:aggregation_coding_table_2} and \ref{tab:aggregation_coding_table_3}). 

\begin{table}[]\small
    \centering
    \begin{tabular}{p{4cm}p{4cm}p{4cm}}
    \toprule
    Core phenomenon (\textbf{bold}) or theme (\textit{italics}) & \textbf{Spectrum} & \textbf{Definition} \\
    \midrule
    \textbf{Relevance of explanation to the overall understanding} 
    & -- Contribution vs limited contribution vs no contribution &
    How the explanation contributes to the understanding of the individual (in general terms and the concrete information).\\
    \textit{Understandability for the average consumer} 
    & -- Difficult vs possible to understand & \multirow{2}{4cm}{The capacity of the average consumer (well-informed and reasonably observant and circumspect individual) to process and understand the explanation/information provided.} \\
    & -- The data controller has the intention to provide explanations that are confusing vs understandable to the data subject. & \\
    &&\\
    \textbf{Appropriateness of the delivery method and format} 
    & -- Confusing vs clear (possibility to be) & \multirow{2}{4cm}{How suitable is the format of the explanation/information provided and the selected methodology.} \\
    & -- Misleading vs helpful (intention of the data controller) & \\
    &&\\
    \textbf{Suitability for exercising rights} 
    & -- No actionability vs limited actionability vs actionability & How appropiate are the explanations/information provided to the individual if they want to act upon the decision affecting them. \\
    &&\\
    \textbf{Significance of the decision's outcome in terms of information requirements} 
    & -- The explanation should be/is adapted in its information w.r.t.~whether the outcome is good or bad for the data subject, vs it should/is not. &	Whether the impact of the decision determines the threshold and amount of information requested by the transparency requirements. \\
    &&\\
    \textbf{Completeness of information} 
    & -- Adequate vs lacking & The amount of information provided. \\
    &&\\
    \textbf{Usefulness and necessity of information based on typology} 
    & -- Request for local vs global vs combination & The requirement and necessity to provide specific and particular information about the decision. \\
    \bottomrule
    \end{tabular}
    \caption{Questionnaire coding table.}
    \label{tab:questionnaire_coding_table_0}
\end{table}

\begin{center}
\begin{table}[]\small
    \centering
    \begin{tabular}{p{3cm}p{3cm}p{6cm}}
    \toprule
    Core phenomenon (\textbf{bold}), theme (\textit{italics}) or sub-theme & \textbf{Category} & \textbf{Spectrum or sub-category (if) and definition} \\
    \midrule
    \textbf{Presented explanations (PE)} & & Explanations presented in the questionnaire and the interview. \\
    \textit{Understanding} & & \textbf{Spectrum}: hard vs. easy \\
    && Whether the explanations were hard or easy to understand. \\
    \textit{Information} & & \textbf{Spectrum}: missing vs. complete \\
    && Whether the information in the explanations was complete or not. \\
    \midrule
    \textbf{High-level topics} & & Sub-themes that do not fall under PE/validity but emerged dynamically and refer to high-level considerations \\
    \textit{Interdisciplinarity} & &  The design of an explanation needs an interdisciplinary perspective. XAI must be closely tied to fairness considerations about the decision. \\
    \textit{Team} & & The design of an explanation is team work, 
    the data subject must be incorporated (e.g., user-centered design approaches). \\
    \textit{Task} & & The task/function of an explanation is subject of discussion. 
    \\
    \bottomrule
    \end{tabular}
    \caption{Aggregation coding table, part 1.}
    \label{tab:aggregation_coding_table_1}
\end{table}
\end{center}

\begin{center}
\begin{table}[]\small
    \centering
    \begin{tabular}{p{3cm}p{3cm}p{6cm}}
    \toprule
    Core phenomenon (\textbf{bold}), theme (\textit{italics}) or sub-theme & \textbf{Category} & \textbf{Spectrum or sub-category (if) and definition} \\
    \midrule
    \textbf{Validity} & & Standard to evaluate the PE according to its applicable techno-legal framework. \\
    \textit{Parties' relationship} & & Data controller and subject are parties in a specific relationship. 
    \\
    {Duties and rights} 
    & \multirow{2}{3cm}{Interests and integrity} & \textbf{Spectrum}: aligned vs conflicted \\
    & & The parties want to achieve something through their relationship, each with their own interests. These  
    can be aligned (both want the same/similar) or conflicted. The interpretation of \enquote{interest} is broad.
    \\
    & & \textbf{Spectrum}: aligned vs conflicted objective/interest w.r.t.~to legal rights and duties \\ 
    & & The parties want to fulfill their duty and enjoy they rights (aligned), in contrast with one of them or both wanting not to/trying to avoid to fulfill their duties/rights (conflicted). We look at the controller and subject separately. \\
    & {Specific Interests} & \textbf{Sub-category}: Happy data subject \\
    & & Centering the data subject (when designing XAI, as part of the perspective of the business or from a protection perspective). \\
    & & \textbf{Sub-category}: Business \\
    & & XAI as a tool that supports the interests of the business, i.e., by generating positive attention.\\
    {Risks}
    & Legal & Exposure to the chance of a legal risk (e.g., 
    data protection breach). \\
    & Technical & Exposure to the chance of a technical risk (e.g. data leakage).
    \\
    {Control mechanism} 
    & Oversight & Proposal of oversight over ADM system and activities of the data controller. \\
    \bottomrule
    \end{tabular}
    \caption{Aggregation coding table, part 2.}
    \label{tab:aggregation_coding_table_2}
\end{table}
\end{center}

\begin{center}
\begin{table}[]\small
    \centering
    \begin{tabular}{p{3cm}p{3cm}p{6cm}}
    \toprule
    Core phenomenon (\textbf{bold}), theme (\textit{italics}) or sub-theme & \textbf{Category} & \textbf{Spectrum or sub-category (if) and definition}\\
    \midrule
    \textbf{Validity} (cont.) & & \\
    \textit{Normativity} & & The way explanations/information ought to be provided according to Article 22 of the GDPR \\
    \multirow{2}{3cm}{Appropriateness of XAI method} 
    & \multirow{2}{3cm}{Global-local methodology} & \textbf{Spectrum}: complementary vs. exclusive \\
    & & Whether global and local explanations should be use combined or exclusively. \\
    \multirow{2}{3cm}{Appropriateness of delivery} 
    & {Format} & \textbf{Spectrum}: complementary vs. exclusive \\
    & & Whether different 
    formats (e.g., plot or text) should be use combined or exclusively. 
    \\
    & Presentation & \textbf{Spectrum}: layered/interactive vs static \\
    & & Whether the explanation should be designed in layers,
    that responds to different needs of the data subject, or integrate an interaction, 
    or whether they should be unique to all data subjects. 
    \\
    & Trade-off & Trade-off between complexity/information-richness versus understandability/intuitiveness.
    \\
    {Adaptability (of explanation to data subject)}
    & Audience & Considering the audience/end-user of an explanation. \\
    & Accessibility & Accessibility of the explanation (e.g., translation into other languages, easy to find online/per letter, easy to digest). \\
    & Domain expertise & \textbf{Spectrum}: Complete vs missing \\
    & & The explanation lacks/provides domain expertise needed to understand the explanation. \\
    {Actionablity (for the data subject)} 
    & Justification & Whether the explanation aims to explain how it respects the norms/laws applicable to it (justificability) or explains how the decision was reached 
    (no/limited justificability). Justificability is considered separate from explainability, i.e., an "add-on". \\
    {Contextual Dependency}
    & Legal & Legal norms that apply to the explanation depend on the context. \\
    & Explanation content & Information the explanation is providing depends on context. \\
    & Explanation method and delivery & The method used and the delivery depend on the context. \\
    \bottomrule
    \end{tabular}
    \caption{Aggregation coding table, part 3.}
    \label{tab:aggregation_coding_table_3}
\end{table}
\end{center}

\section{Multiple choice question results}
\label{sec:appendix_mc_questions}

We display remaining answers to the multiple choice questions in Figure \ref{fig:mc_1}.

\renewcommand{\thefigure}{8}
\begin{figure}
    \centering
    \includegraphics[width=.45\textwidth]{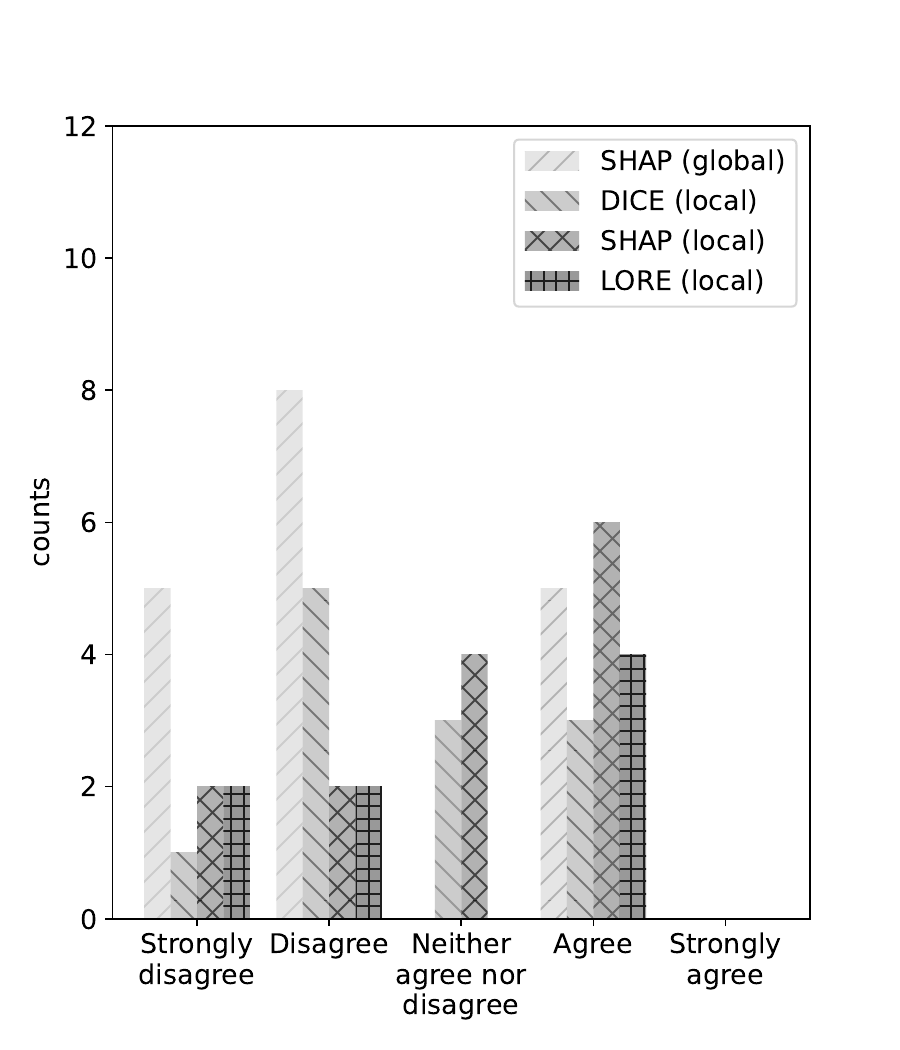}
    \includegraphics[width=.45\textwidth]{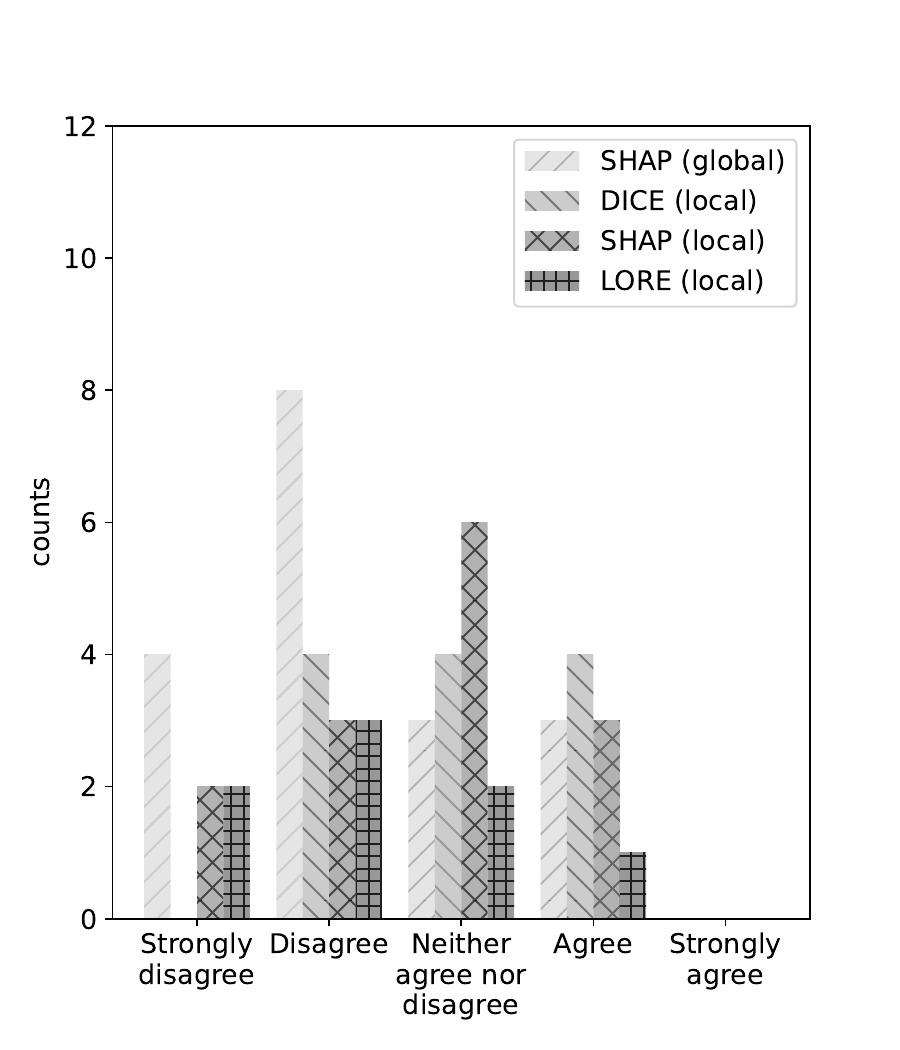}
    \includegraphics[width=.45\textwidth]{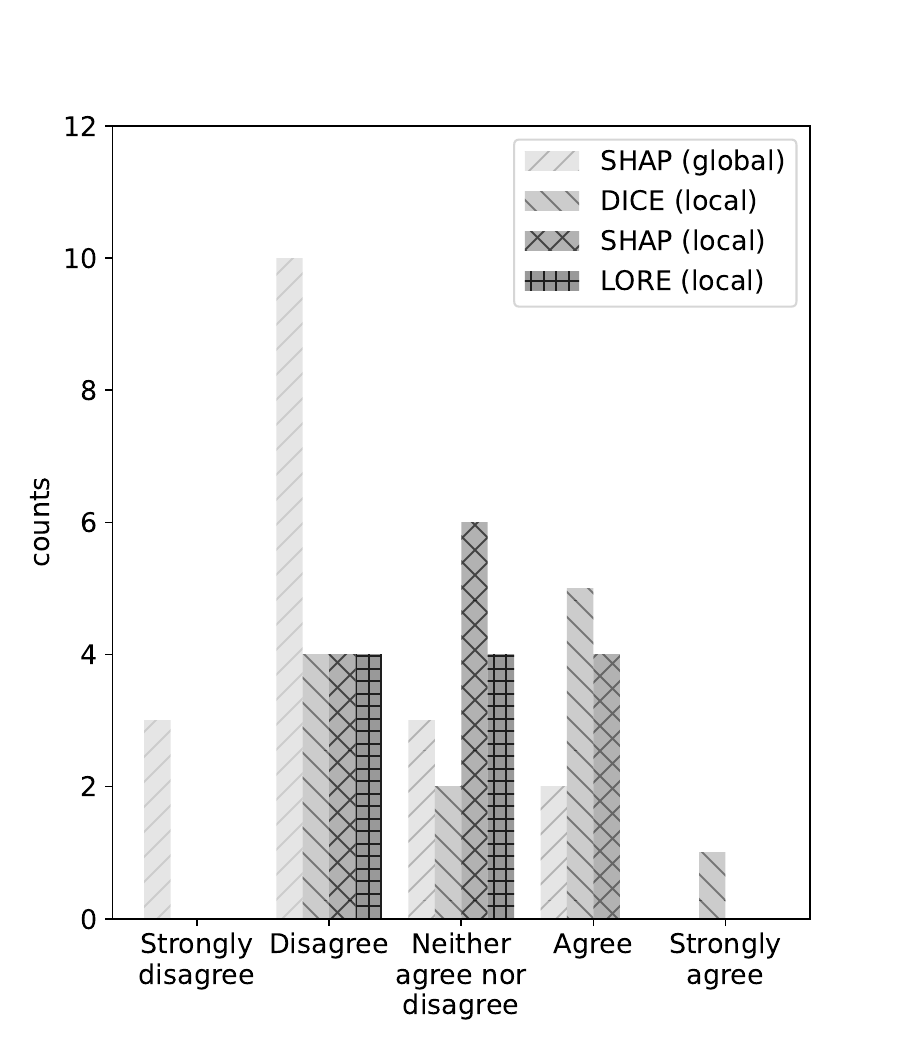}
    \includegraphics[width=.45\textwidth]{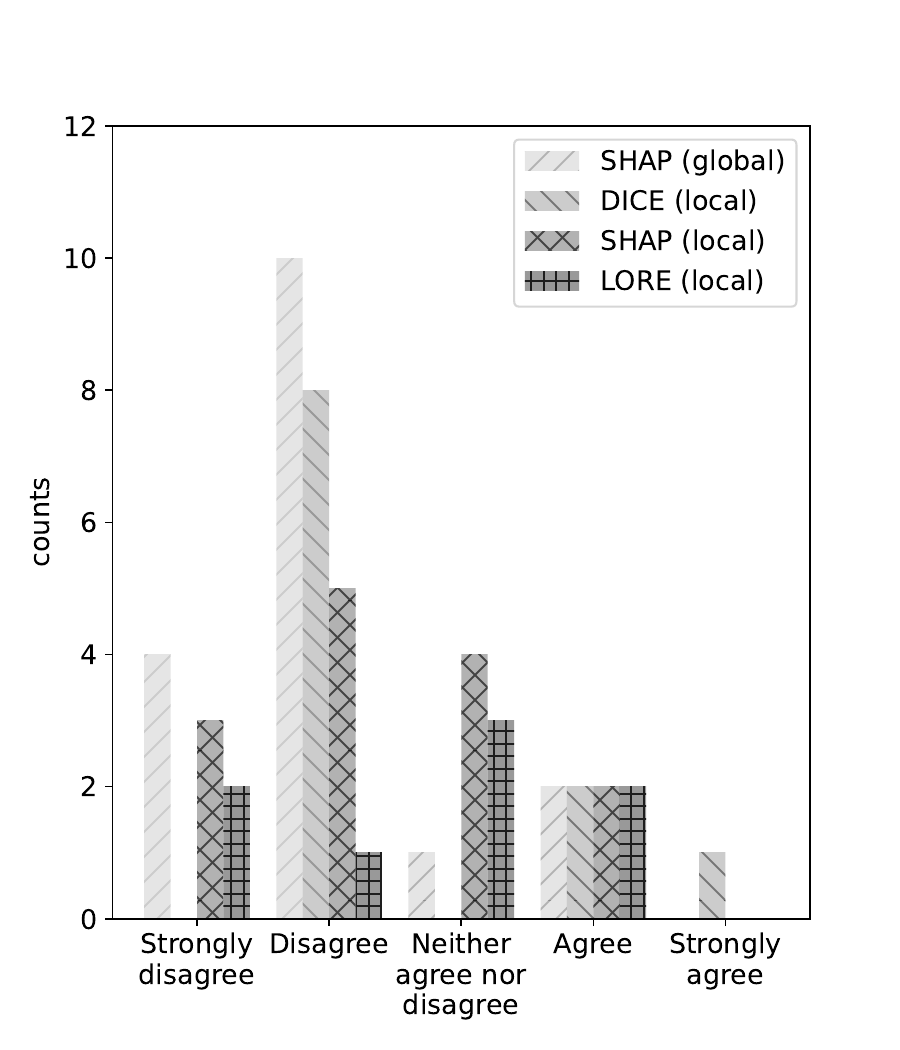}
    \caption{Answers to multiple choice questions. Upper left: answering to \enquote{The explanation presented was clear and understandable for an “average individual”}, upper right: answering to \enquote{The explanation presented provides sufficient information about the decision to ensure that the person affected by the decision understands the reasons and motives behind it}, lower left: answering to \enquote{The explanation presented allows the data subject to effectively exercise her right to contest such automated decisions [if deemed appropriate], in accordance with the safeguards established in Article 22 paragraphs 3}, lower right: answering to \enquote{The explanation provided allows the data subject to verify the lawfulness and fairness of the automated decision in regard to other sectoral laws applicable to the case (e.g. consumer law, contract law, anti-discrimination law)}.}
    \label{fig:mc_1}
\end{figure}

\end{appendices}

\end{document}